\documentclass{aa}  

\usepackage{graphicx}
\usepackage{txfonts}
\usepackage{booktabs}
 \usepackage{hyperref}

\usepackage{xcolor}

\begin{document}

   \title{Dynamical friction and massive black hole orbits: analytical predictions and numerical solutions}

   \author{Alice Damiano
          \inst{1,2,3,4,5},
          Stefano Borgani \inst{1,2,3,4,5},
          Milena Valentini \inst{1,2,3,4,5}, 
          Giuseppe Murante \inst{2}, 
          Luca Tornatore \inst{2,5}, \\ 
          Petr Strakos 
          \inst{6},
          \and
          Milan Jaros 
          \inst{6}}
          

  \institute{Dipartimento di Fisica dell'Universit\`a di Trieste, Sez. di Astronomia, via Tiepolo 11, I-34131 Trieste, Italy    
     \and
      INAF -- Osservatorio Astronomico di Trieste, via Tiepolo 11, I-34131, Trieste, Italy 
        \and
        IFPU, Institute for Fundamental Physics of the Universe, Via Beirut 2, 34014 Trieste, Italy
        \and
        INFN, Instituto Nazionale di Fisica Nucleare, Via Valerio 2, I-34127, Trieste, Italy
        \and
        ICSC - Italian Research Center on High Performance Computing, Big Data and Quantum Computing, via Magnanelli 2, 40033, Casalecchio di Reno, Italy
        \and
        IT4Innovations, VSB – Technical University of Ostrava, 17. listopadu 2172/15, 708 00 Ostrava-Poruba, Czech Republic}

    \titlerunning{Dynamical friction and massive BH orbits}
   \authorrunning{Alice Damiano et al.}

   \date{Received ----}

\abstract{} 
   {We investigate the orbital decay of a massive BH embedded in a dark matter halo and a stellar bulge, using both analytical and numerical simulations with the aim of developing and validating a reliable dynamical friction (DF) correction across simulation resolutions.} 
{ We develop a Python-based library to solve the equations of motion of the BH and provide an analytical framework for the numerical results. Then, we carry out simulations at different resolutions and for different softening choices using the Tree-PM code \texttt{OpenGADGET3}, where we implement an improved DF correction based on a kernel-weighted local density estimation. }
  {Our results demonstrate that the DF correction significantly accelerates BH sinking and ensures convergence with increasing resolution, closely matching analytical predictions. We find that in low-resolution regimes—particularly when the BH mass is smaller than that of the background particles—our DF model still effectively controls BH dynamics. Contrary to expectations, the inclusion of a stellar bulge can delay sinking due to numerical heating, an effect partially mitigated by the DF correction. } 
  {We conclude that our refined DF implementation provides a robust framework for modeling BH dynamics both in controlled simulation setups of galaxies and in large-scale cosmological simulations. This will be crucial for future simulation campaigns, to enable more accurate predictions of AGN accretion and feedback, and to estimate gravitational-wave event rates.}


   \maketitle
%

\section{Introduction}

Massive black holes (MBHs) reside at the center of massive galaxies \citep{kormendy1995inward, ferrarese_fundamental_2000-1}, and they gradually sink toward the core of a newly-formed galaxy when their host galaxies undergo mergers \citep{callegari2009pairing, mayer2007rapid, capelo_growth_2015, comerford2009inspiralling, volonteri2015massive}, eventually forming a dual or binary system  \citep{volonteri2022dual, dotti2010dual, volonteri2021origins, cao2025dual, merritt2004massive}. 
The diversity in MBH population demography arises from the dynamical processes governing their motion. In the dense stellar environments of galactic centers, MBHs experience dynamical friction (DF) \citep{chandrasekhar}, a drag force exerted by the surrounding stellar and DM distribution. 

DF not only anchors MBHs to the galactic center but also facilitates their gradual loss of angular momentum, driving their inspiral toward the core \citep{antonini2011dynamical, li2020pairing, vecchio1994orbital, volonteri_cosmic_2016, Just_2010, boylan2008dynamical}.
Determining the timescale for MBH sinking and coalescence remains a complex issue stemming from  the complex dynamical environment in which MBHs evolve—a regime that deviates significantly from idealized analytical descriptions and demands both high numerical precision and spatial resolution in simulations to be properly captured.

First introduced by \cite{chandrasekhar}, the DF force describes the gravitational drag exerted by a background medium of stars or dark matter (DM) on a massive object moving through it. In the simplified case where the surrounding particles are negligible in mass compared to the perturber, and are distributed in a uniform and homogeneous medium, a simple analytical formula for the sinking timescale can be derived (\citealt{binney_galactic_1987}).
However, it is unlikely that the BH experiences the passage through a homogeneous and isotropic medium when infalling to the galactic core. 
Furthermore, the choice of the maximum radius where to account for the DF is a free parameter whose choice has a significant impact on the resulting timescale for sinking. 

While the analytical solution derived from the Chandrasekhar DF formula can be inaccurate, searching for a numerical solution is equally challenging (\citealt{taffoni_life_2003-1}). The primary difficulty lies in the large spatial range required to properly simulate the infall of a MBH toward the galactic center, alongside the high spatial and mass resolution necessary to accurately capture gravitational interactions, including both DF and large-scale galactic torques (\citealt{bortolas2020global}).
In the context of N-body simulations, both coarse mass resolution and too aggressive gravitational force softening can introduce numerical heating into the BH’s motion. 

To mitigate this limitation, several ad hoc prescriptions have been proposed. Among them, a widely used approach involves repositioning the BH at the local minimum of the gravitational potential (\citealt{springeldimatteo2005, dave_simba_2019, bassini_black_2019}). However, this method suffers from significant drawbacks, particularly in environments having a  shallow potential well or during galaxy mergers, where it may lead to unphysical BH trajectories (e.g., \citealt{damiano_dynamical_2024} and references therein). Another serious limitation in correctly following orbits of BH particles arises close to epoch of BH seeding. In this regime BH masses are generally smaller than that of surrounding DM and stellar particles, thus rendering DF poorly described by the N-body solver. To mitigate this limitation, it has been proposed to artificially boost the BH "dynamical mass", i.e. the mass felt by gravity, so as to enhance the DF effect \citep[e.g.][]{Curtis_2015}.

Currently, a variety of DF-based sub-resolution models for BH dynamics are implemented across different cosmological and galaxy formation simulations, each attempting to compensate for numerical limitations while ensuring realistic BH sinking and merging processes \citep{hirschmann_cosmological_2014, tremmel_off_2015, chen, ma_new_2023}.
All such models involve the addition of corrective terms to the BH acceleration predicted by the N-body solver, to account for the DF exerted by the surrounding particles. 
However, it is generally accepted that even with a DF correction, the dynamics of MBHs remains unresolved when the BH mass is lower than approximately twice the mass of the surrounding particles \citep{genina_calibrated_2024}. In such cases, numerical inaccuracies can still dominate the BH's trajectory, making it challenging to accurately describe the expected sinking process. A validation of these numerical techniques requires testing them with controlled and idealized numerical experiments, which satisfy the assumptions on which analytical predictions are based. 

In this paper, we investigate the infall of a BH in a DM halo, with and without the inclusion of a stellar bulge,  to address the following key questions: how do the sinking timescales of the BH compare with and without the DF correction? Is there numerical convergence  of the sinking timescale at increasing resolution when using the DF correction? How do numerical results compare with analytical predictions at progressively increasing resolution and including a DF correction? How does the presence of a stellar bulge impact the BH dynamics? Finally, can a DF correction be developed to accurately control the BH dynamics in the critical regime where the BH mass is smaller than that of the surrounding particles?
To address these questions, we introduce a refined DF prescription, built upon the framework of \cite{damiano_dynamical_2024}, but tailored to accurately model the challenging regime where the BH mass is smaller than that of the surrounding particles. This new approach extends the validity of our treatment of sub-resolution DF correction across a wider range of mass ratios, marking a significant improvement in simulating BH dynamics, especially in resolution regimes typical of cosmological simulations. Additionally, we introduce the Orbital TImescale for Sinking (OTIS) library, an open-source Python library designed to solve the equations of motion of a BH infalling into a Navarro-Frenk-White (NFW) dark matter halo \citep{navarro_structure_1996, navarro_universal_1997}, complemented with a stellar bulge. This library provides a fast and flexible tool to compare the numerical results with theoretical expectations. Equipped with these tools, we consider the infall of a BH in an NFW DM halo, possibly also including a stellar bulge following a \cite{hernquist_analytical_1990} density profile. We compare the BH sinking timescales against the analytical predictions across different resolutions, with and without the DF correction, focusing in particular on the challenging regime when the BH is less massive than the surrounding particles.

This paper is organized as follows:  Sect. \ref{Analytical} and Sect. \ref{Numerical}  describe the methodology for the analytical framework and the numerical setup, respectively. Sect. \ref{Sec:Results} presents our results for BH infall in a DM halo (Sect. \ref{Results.DMhalo}) and when including a stellar bulge (Sect. \ref{Results.Stellarbulge}). We pay particular attention to cases when the BH is less massive than its surrounding particles, discussed in Sects. \ref{DM.lowres} and \ref{Stellar.lowres}. Finally, Sect. \ref{Discussion} discusses our key results, while Sect. \ref{conclusions} presents our main conclusions.



\section{Analytical framework} \label{Analytical}
In this section, we outline the steps required to derive the analytical solution for the sinking timescale of a BH moving within an NFW potential (Sect. \ref{subsect: analytical_circular}, \ref{subect:elliptical_analytics}), later coupled with a Hernquist stellar bulge (Sect. \ref{subsect: stellarbulge}), and subject to DF  from the surrounding DM and stellar particles.  Finally, Sect. \ref{subsect:otis} introduces the python library we developed to integrate the equations of motion of the BH in these scenarios.

\subsection{Infalling BH on a circular orbit in an NFW halo} \label{subsect: analytical_circular}

The NFW halo density profile $\rm \rho_H$ can be expressed as a function of the halo-centric distance $\rm r$ as (\citealt{navarro_structure_1996, navarro_universal_1997}): 
\begin{equation} \label{rho_nfw}
  \rm \rho_H (r) = \frac{\rho_s}{{\frac{r}{r_s}\left(1+\frac{r}{r_s}\right)^2}} 
\end{equation}
where $\rm r_{s}$ is the scale radius, defined by the virial radius $\rm r_{vir}$ \footnote{Using the same formalism as \cite{springel_1999}, we define the  virial radius as the radius within which the mean halo density corresponds to an overdensity $\delta_c$= 200 times the critical cosmic density at redshift z, i.e. $\rm \rho_c(z)=3H(z)/8 \pi G$.}
 and the concentration parameter $c$ such that $\rm r_s = r_{vir}/c $, and $\rm \rho_s$ is the mass density at $\rm r_{vir}$. 
By integrating Eq.\eqref{rho_nfw} we obtain the expression for the halo mass enclosed within a radius r:
\begin{equation} \label{mass_nfw}
    \rm M_{H}(r)=4 \pi \rho_s r_s^3 \left[ \ln \left(1+\frac{r}{r_s}\right)-\frac{r}{(r_s+r)} \right].
\end{equation}
Assuming to place a BH at a given distance $\rm r$ from the center of the halo on a circular orbit, its initial circular velocity will be 
\begin{equation} \label{velocity_circ}
    \rm  v_{c} (r)= \sqrt{ \rm \frac{GM_{H}(r)}{r}},
\end{equation}
whereas the gravitational potential at the BH position is obtained through the Poisson equation:
\begin{equation}
   \rm  \Phi_{BH}(r) = -4 \pi G \rho_s \frac{r_s^3}{r} ln\left( 1+ \frac{r}{r_s} \right).
\end{equation}
Therefore, the radial acceleration felt by the BH takes the expression:
\begin{equation} \label{accnfw}
 \rm \vec{ \rm a}_{BH}(\vec{r}) = -\vec{\nabla} \Phi_{BH} = G \frac{M_{vir}}{ln(1+c)-\frac{c}{1+c}} \frac{1}{r^2} \left[ \frac{r}{r_s+r} - ln \left(1+\frac{r}{r_s} \right)\right]\hat{r},
\end{equation}
where $\rm M_{vir}$ is the halo mass within $\rm r=r_{vir}$ obtained from Eq.\eqref{mass_nfw} and the "hat" indicates the versor of a vector.
When the halo is sampled with a finite number of particles of mass $\rm m_p$, they exert a friction onto the BH travelling through them. This force causes a gradual loss in the angular momentum of the BH and from simple dynamical arguments we can analytically derive the path of the infalling BH as it moves inwards toward the center. \cite{rodriguez_new_2018} derived the equation for the orbital evolution of the BH for circular orbits around the center of the halo, following a \cite{hernquist_analytical_1990} profile. We employ the same methodology for an NFW density profile of the halo, assuming that the BH is initially placed on a circular orbit around the halo center. 
In Sect. \ref{subect:elliptical_analytics} we will relax this assumption, by considering the possibility of an initial ellipticity of the BH orbit.

The magnitude of the torque acting on the infalling BH of mass $\rm m_{BH}$ when orbiting toward the center takes the general form:
\begin{eqnarray}
    \rm \frac{dL}{dt}&=& \rm m_{BH}\left( \rm v_{c}+r \frac{ \rm dv_c(r)}{ \rm dr} \right) \frac{ \rm dr}{ \rm dt}\\
    &=& \rm m_{BH} \frac{v_c(r)}{2} \left( 1+ \frac{d \log M_H(r) }{d \log\  r }\right) \frac{dr}{dt} \,.
\end{eqnarray}
By replacing the NFW mass profile defined in Eq.\eqref{mass_nfw} and its derivative in the previous equation, we obtain:
\begin{eqnarray} \label{finaldLdt}
     \rm \frac{d  L}{dt} &=& \rm  m_{BH}\frac{v_c(r)}{2} \left\{1+ \frac{r^2}{r_s^2 \left( 1+ \frac{r}{r_s}\right)^2 \left[ ln \left(1+\frac{r}{r_s}\right) - \frac{r}{r_s+r} \right]}\right\} \frac{d r}{dt}\\
     &=& \rm m_{BH} \mathcal{F}(v_c, r_s, r) \frac{dr}{dt} \,,
\end{eqnarray}
where we introduce the function: 
\begin{equation}
\rm  \mathcal{F}(v_c, r_s, r) =  \rm  \frac{v_c(r)}{2} \left\{1+ \frac{r^2}{r_s^2 \left( 1+ \frac{r}{r_s}\right)^2 \left[ ln \left(1+\frac{r}{r_s}\right) - \frac{r}{r_s+r} \right]}\right\}.
\end{equation}
The angular momentum torque on a BH moving on a circular orbit is caused by the DF force, which is perpendicular to the radial direction. Therefore, the torque magnitude can be expressed as  (\citealt{rodriguez_new_2018}):
\begin{equation} \label{LlossDF}
\rm    \frac{d L}{dt}= m_{BH} a_{DF} r \,,
\end{equation}
where we introduced the DF acceleration $\rm a _{DF}$ acting onto the BH. Under specific assumptions -- i.e., $(i)$ the surrounding particle distribution must be infinite, isotropic and homogeneous (thus ensuring that the perpendicular component of the velocity change during two-body scattering is negligible), $(ii)$ self-interactions among the surrounding particles are negligible and $(iii)$ the mass of these particles is much smaller than the BH mass, i.e.  $ \rm m_p \ll m_{BH}$ -- this acceleration is given by 
(\citealt{binney_galactic_1987}):
\begin{equation} \label{general.df}
    \rm\vec{\rm a}_{\rm DF} (r) = \rm -8  \pi^2 \log(1+\Lambda^2)\,  G^2 m_{p}(m_{BH}+m_p) \frac{ \int_0^{v_{BH}} v^2 f(v) dv}{v_{BH}^2}\, \hat{v}_{BH},
\end{equation}
where $ \rm v_{BH}$ is the BH velocity, G is the gravitational constant, $\rm f(v)$ is the velocity distribution function of the surrounding particles and $\rm \Lambda = b_{max}/b_{min}$ where $\rm b_{max}, b_{min}$ are the maximum and minimum impact parameter to account for. 
Assuming that the surrounding particle distribution is Maxwellian and $\rm \Lambda$ is large so that $\rm log(1+\Lambda^2) \sim 1/2 \log(\Lambda)$, \cite{binney_galactic_1987} showed that this expression can be simplified as:
\begin{equation} \label{adf}
     \rm \vec{\rm a}_{DF} (r) =  \rm - \frac{4 \pi \ln (\Lambda) G^2 \rho (m_{BH}+m_p)}{v_{BH}^2} \left[ erf( \mathcal{X}) - \frac{2 \mathcal{X}}{\sqrt{\pi}} e^{-\mathcal{X}^2}\right] \hat{v}_{BH}\,.
\end{equation}
Here $\rm erf$ is the error function, $\rho$ is the density surrounding the BH, while $\mathcal{X}$ depends on the velocity of the BH $\rm v_{BH}$,  and on the velocity dispersion of the surrounding medium $\sigma$ as: 
\begin{equation} \label{chi}
    \mathcal{X}=  \rm \frac{v_{BH}}{\sqrt{2} \sigma}.
\end{equation}
Combining Eq. \eqref{finaldLdt}, \eqref{LlossDF} and \eqref{adf}, we obtain an expression for the radial velocity of the BH particle inspiralling into a NFW halo:
\begin{equation} \label{originaldrdt}
   \rm   \frac{d\vec{\rm r}}{dt} =  \rm - \frac{4 \pi \ln (\Lambda) G^2 \rho_H (M_{BH}+m_p)}{v_{BH}^2 \mathcal{F}(v_c, r_s, r)} \left[ erf( \mathcal{X}) - \frac{2 \mathcal{X}}{\sqrt{\pi}} e^{-\mathcal{X}^2}\right]  r \hat{v}_{BH}.
\end{equation}
Lastly, we derive the velocity dispersion appearing in Eq.\eqref{chi} from the Jeans equation: 
\begin{equation} \label{vel.disp.halo}
   \rm  \sigma^2_{{H}}(r) = \frac{1}{\rho_H(r)} \int_r^{\infty} \rho_H(r') \frac{GM_H(r')}{r'^2} dr'
\end{equation}
that can be integrated using Eq. \eqref{rho_nfw} and \eqref{mass_nfw}.

\subsubsection{Elliptical orbits} \label{subect:elliptical_analytics}
The previous discussion related to an initially circular orbit can be extended to elliptical orbits. In this case, the assumption that the orbits remains circular while the BH is sinking is clearly no longer valid. Therefore, Eq.\eqref{LlossDF} does not hold. To derive an analytical expression of the sinking timescale in this more general context, we decompose the total BH acceleration into a first component which is instantaneously directed toward the center of the NFW halo and a second component which is always antiparallel to the BH velocity vector: 
\begin{equation} \label{acc_ecc}
\rm a_{BH}= -a_{H}(r) \hat{ r} -a_{DF} \hat{v}_{\rm BH}.
\end{equation}
The first term in the above equation is given by Eq.\eqref{accnfw}, while the second one by Eq.\eqref{originaldrdt}. 
In the following, we refer to the orbital eccentricity using the standard definition: 
\begin{equation} \label{eccentricity}
    e  = \rm  \sqrt{ \rm 1-\frac{L^2}{GM_{H}m_{BH}^2 d_a}} \,,
\end{equation}
where $\rm L$ is the angular momentum of the BH and $ \rm d_a$ is the apocentric distance.

\subsection{Adding a stellar bulge} \label{subsect: stellarbulge}

 Following the same formalism as  \cite{springel_1999}, we model the stellar bulge, that we superimpose to the NFW DM halo, as a non-rotating spheroid following a Hernquist density profile: 
\begin{equation} \label{density.bulge}
 \rm \rho_b(r)=\frac{M_B}{2 \pi}\frac{r_b}{r(r_b+r)^3}
\end{equation}
where $\rm M_B$ is the total bulge mass and $r_b$ its scale radius. 
Integrating over the radial coordinate we  then obtain the bulge mass as a function of the radius:
\begin{equation}\label{mass.bulge}
 \rm M_b(r)=\frac{M_Br^2}{(r_b+r)^2}\,.
\end{equation}
The gravitational potential associated to this bulge component is given by:
\begin{equation}
    \rm \Phi_b = -\frac{GM_B}{r+r_b},
\end{equation}
so that the acceleration acting on a BH infalling into the bulge is: 
\begin{equation} \label{acc_bulge}
 \rm    \vec{\rm a}_{BH}= -\vec{\nabla} \Phi_B = -\frac{GM_B}{(r+r_b)^2} \hat{r}.
\end{equation}
Solving the Jeans equation, we can derive the velocity dispersion for the bulge component:

\begin{multline} \label{vel.disp.bulge}
\sigma_b^2(r) = \rm -G m_b r (r_b + r)^3 \left[ 
\frac{\ln |r| - \ln |r + r_b|}{r_b^5} + \frac{1}{r_b^4 (r + r_b)} + \right. \\
\rm \left. + \frac{1}{2 r_b^3 (r + r_b)^2} + \frac{1}{3 r_b^2 (r + r_b)^3} 
+ \frac{1}{4 r_b (r + r_b)^4} \right] \,.
\end{multline}
When adding the bulge counterpart, the resulting density and velocity distribution of the system can be expressed as the sum of the halo and the bulge components. The circular velocity at radius $r$ of the BH in the composite system is then given by: 
\begin{equation}
  \rm   v_{c,BH} (r)= \rm \sqrt{\frac{ \rm G\left[M_H(r)+M_b(r)\right]}{\rm r}}.
\end{equation}

\subsection{Numerical integration: the Orbital TImescale for Sinking library} \label{subsect:otis}
To apply the analytical framework presented in the previous sections, we developed the Orbital TImescale for Sinking (OTIS) Python library,  designed to integrate the BH equations of motion in an NFW halo, eventually including also a stellar core at its center. The code is publicy available at: \url{https://github.com/alicedamiano5/OTIS.git }. OTIS enables the computation of the orbital sinking timescale by solving the equations of motion of the BH moving under the influence of gravitational attraction and DF. 
The gravitational acceleration is computed self-consistently from the NFW and bulge potential, while the DF acceleration is modeled according to Eq.\eqref{general.df}. The modular structure of the code enables a flexible configuration of interpolation techniques and can account for different initial conditions for both the halo and the BH, making it well-suited for fast calculations of sinking timescales in both idealized and simple multi-component systems.

\begin{figure}
    \centering
    \includegraphics[width=0.9\linewidth]{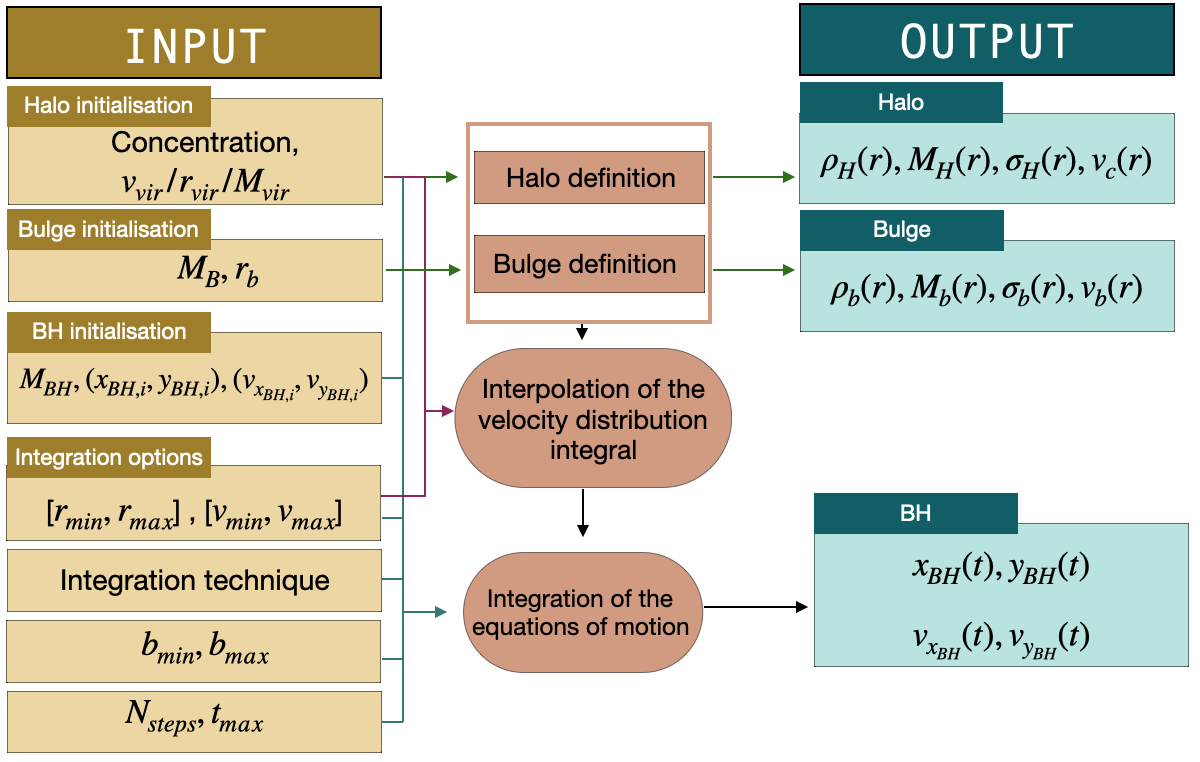}
    \caption{Schematic flowchart of the OTIS library. The code workflow starts with the initialization of the DM halo parameters, including the concentration parameter and virial quantities, eventually followed by the initialization of the stellar Hernquist bulge. The density, mass, and velocity dispersion profiles of both the halo and the bulge can be hence retrieved. The BH is then initialized with its mass and phase-space coordinates. The integral of the velocity distribution function is precomputed on a phase-space grid, enabling fast interpolation during the numerical integration of the equations of motion. User-configurable options allow customization of the interpolation technique, integration method, time steps, and impact parameters.}
    \label{fig:otis_code_flow}
\end{figure}

Figure \ref{fig:otis_code_flow} summarizes the OTIS code flow. Given the halo concentration parameter and one of the virial quantities (radius, velocity or mass), the code computes the DM density, mass, and velocity dispersion profiles, assuming an input value for the Hubble parameter. When also the bulge component is included, its initialization requires the definition of the bulge total mass and scale radius. The BH is initialized at a user-defined phase-space coordinate $ \rm ({\rm x}_{{BH},i}, {\rm y}_{{BH},i},{\rm v}_{x_{BH},i},{\rm v}_{y_{BH},i} )$ with mass $\rm M_{BH}$.The trajectory of the BH is assumed to be on a plane intersecting the halo center, reducing the problem to a two-dimensional one. 

In our implementation, the velocity dispersion varies with the radius, as described by Eq. \eqref{vel.disp.halo} and \eqref{vel.disp.bulge}. Consequently, the integral entering Eq.\eqref{general.df} depends itself on the radial coordinate.
To optimize performance, OTIS precomputes this integral on a user-configurable phase-space grid, which is then used during the integration of the equations of motion.
 The library provides various user-configurable options, including the interpolation method (from the predefined SciPy techniques, \citealt{SciPy-NMeth}), the number of time steps, the integration time limit, and the values of the minimum and maximum impact parameters $ \rm b_{min}$ and $ \rm b_{max}$.

In line with \cite{genina_calibrated_2024} and following the work by \cite{Just_2010}, we assume the following expression for the minimum impact parameter:
\begin{equation} \label{bmin}
 \rm b_{min}= \frac{GM_{BH}}{v_{BH}^2+ 2/3 \sigma^2}\,,
\end{equation}
corresponding to the minimum impact parameter for a $\rm 90°$-deflection two-body encounter occurring at a typical velocity $\rm v_{typ} = v^2+2/3 \sigma^2$ for a Maxwellian velocity distribution function.

\begin{figure*}
    \centering
    \includegraphics[width=1\linewidth]{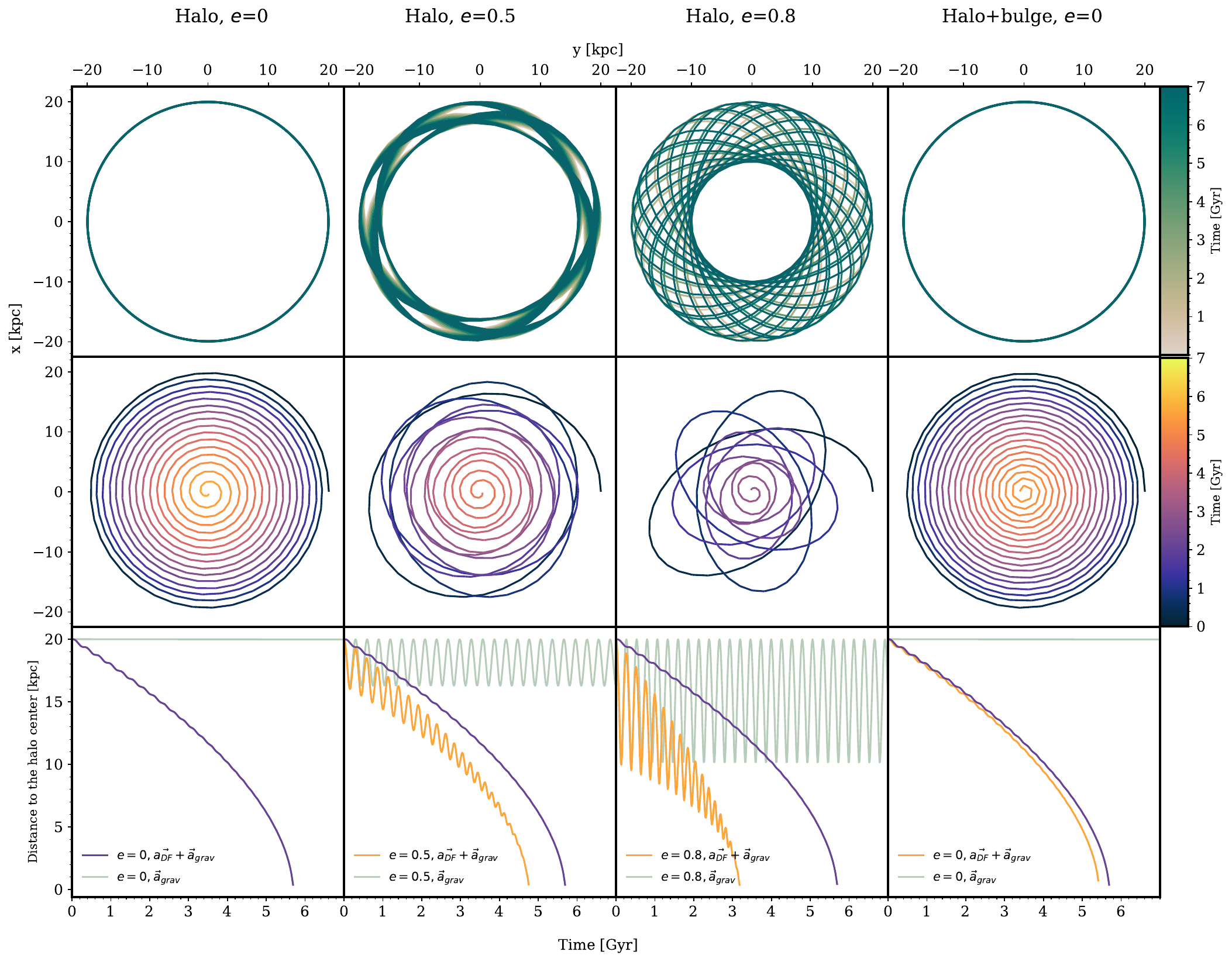}
    \caption{Evolution of a BH embedded in an NFW DM halo, with and without a central stellar bulge, as simulated using the OTIS code. The first three columns correspond to BH trajectories in an NFW halo with initial distance of $20$ kpc from the halo center and orbit ellipticities $e=0$, $e=0.5$, $e=0.8$, while the fourth column includes a central bulge and shows the trajectory of the BH initially at $20$ kpc and placed on a circular orbit. In the first and second row, the trajectory color-code refers to the time elapsed since the beginning of the simulation, as indicated by the color map.
Top row: BH trajectories when it is embedded within the static gravitational potential generated by the mass distribution, thus neglecting collisional effects from two-body encounters. Middle row: BH trajectories when collisional effects are included through the analytical description of the DF force (see Eq. \eqref{general.df}). Bottom row: time evolution of the BH’s radial distance from the halo center. The green line shows the distance when $\rm \vec{a}_{BH}$ follows eq. \eqref{accnfw} (or eq. \eqref{acc_bulge} in the last column) mimicing the effect of a BH embedded in a collisionless medium. The purple line adds the DF contribution given by eq. \eqref{general.df} for eccentricity $e=0$ while the distance corresponding to the middle row of each panel is displayed by the orange curve.}
    \label{fig:otis_fig}
\end{figure*}

Figure \ref{fig:otis_fig} shows the outcome of the OTIS code for a BH embedded in an NFW DM halo (first three columns from the left) as well as in a configuration that includes a central stellar bulge having a Hernquist profile  (fourth column). The NFW has total mass $\rm M=10^{13} \ M_\odot$ and concentration $c=4.38$, while the bulge mass is $\rm 10^{11} \ M_\odot$ and its scale radius is $\rm r_s = 7.2$ kpc.  Starting from a BH located in  $(20$ kpc $,0)$, from left to right we show the trajectory of a BH when embedded in an NFW halo and provided with an initially ellipticity $e=0$, $e=0.5$, $e=0.8$, while the fourth column shows the results adding a central bulge and assuming an initial ellipticity $e=0$.

Panels in the first row display BH orbits in a static potential generated by the system, without accounting for any interactions between particles. In this case, the BH acceleration is determined solely by Eq. \eqref{accnfw} and Eq.\eqref{acc_bulge}. Panels in the second row assumes that the potential arises from a system of discrete particles and includes interactions by adding a DF force, as described by Eq.\eqref{adf}. The third row presents the BH radial distance from the halo center as a function of time, comparing three cases: when the DF is absent (green line), the infall when the BH is placed into the DM halo on initial circular orbit ($e=0$, purple line) and the specific orbital configuration shown in the second row (orange line). When DF is included, the BH progressively spirals toward the halo center, whereas in its absence, it remains on a stable circular orbit. For eccentric orbits, the apoapsis-periapsis distance is not constant; instead, DF gradually circularizes the orbit, a well-known effect due to DF (e.g., \citealt{bonetti_dynamical_2020}). The inclusion of the bulge component slightly reduces the sinking timescale, shortening it by approximately 200 Myr.

\section{Numerical simulations} \label{Numerical}

We carried out several simulations with the aim to assess the performances of DF correction at different resolutions, both with and without DF prescriptions to improve the numerical description of BH dynamics. In this section, we describe the specific model that we follow to correct for unresolved DF  (Sect. \ref{subsect:numerical_df}) and the general simulation setup we adopt in this work (Sect. \ref{subsect:simu_setup}). We finally introduce the list of simulations in Sect. \ref{subsect:list_of_simu} whose results will be presented in Sect. \ref{Sec:Results}.

\subsection{ Numerical DF correction}\label{subsect:numerical_df}
Within the context of N-body simulation codes, the gravity solver already accounts for the DF force, at least within the limits allowed by finite force and mass resolution. However, due to this limited resolution, a correction term needs to be introduced  (e.g., \citealt{hirschmann_cosmological_2014,tremmel_off_2015}). Accordingly, the total acceleration felt by a BH particles can be expressed as the sum of the contribution computed by the gravity solver, $\rm \vec{a}_{grav}$, and a correction term $ \rm \vec{a}_{DF}$, due to the unresolved dynamical friction:
\begin{equation}
    \rm \vec{a}_{BH}= \vec{a}_{grav} + \vec{a}_{DF}.
\end{equation}
Throughout this paper, we adopt and refine the model for DF correction introduced by \cite[][hereafter D24]{damiano_dynamical_2024}. Following \cite{tremmel_off_2015}, to avoid double-counting the DF contribution already captured by the gravitational solver, we apply the DF correction only at distances smaller than a chosen maximum impact parameter, denoted as $\rm b_{\text{max,c}}$. In line with this, D24 adopt $\rm b_{\text{max,c}} = \epsilon_{\rm BH}$, where $\epsilon_{\rm BH}$ is the gravitational softening length of the BH particle. In this work, we adopt the same assumption when the BH mass exceeds that of the surrounding DM particles. However, as the BH becomes less massive, we explore alternative values for $\rm b_{\text{max,c}}$ to assess their impact (see Sect.~\ref{DM.lowres}).

Under the assumption that the phase-space density below $ \rm b_{max, c}$ can be approximated as a sum of the discrete contributions from neighbour particles (D24), the DF correction becomes:
\begin{eqnarray}
\label{num.general.df}
    \rm \vec{a}_{DF}= \rm  \sum_j^{N}-2 \pi G m_j (M_{BH}+ m_j)\,{\tilde n_j} \nonumber\\ \times \rm  \ln\left[1+\Lambda(m_j)^2\right] \frac{\vec{v}_{m_j}-\vec{v}_{BH}}{|v_{m_j}-v_{BH}|^3}\,.
\end{eqnarray}
Here $ \rm m_j$ is the mass of the $j$-th particle, $\tilde n_j$ is the local number density  and the summation runs over all the N particles that are found within a distance $ \rm b_{\rm max, c}$ from the target BH particle. For completeness, the expression for $\Lambda(m_j)$, consistent with the definition adopted in D24, is reported in Appendix B, Eq.\eqref{lambda.numerical}.

Assuming constant density within a sphere of radius $b_{\rm max,c}=\epsilon_{BH}$, Eq.\eqref{general.df} reduces to the original formulation of D24 with $\tilde n_j= 3/(4\pi \epsilon_{BH}^3)$:
 \begin{equation} \label{d24}
 \rm \vec{a}_{DF}  =   \frac{3}{2 {\epsilon_{BH}}^3} \sum_{i=j}^{ \rm N(<\epsilon_{BH})} \ln\left[{1+{\Lambda} (m_j)^2}\right]  m_j ({\rm m_{BH}}+m_j) \frac{(\vec{v}_{m,j}-\vec{v}_{\rm BH})}{|\vec{v}_{m,j}-\vec{v}_{ \rm BH}|^3}\,.
\end{equation}
We note that Eq.\eqref{num.general.df} generalizes the formulation of D24 to accommodate different particle masses and non-uniform density around a BH particle. The value and numerical computation of the local number density depends on the specific kernel adopted. After introducing the \texttt{OpenGADGET3} code used in this work  (Sect. \ref{subsect:simu_setup}), we detail the calculation of the number density in Sect. \ref{DM.lowres}.

Interestingly, the DF correction introduced in this work can be formally derived from the first-order diffusion coefficient in the Fokker-Planck equation, derived from the collisional Boltzmann equation to describe the evolution of the phase-space probability density function  of a self-gravitating fluid. This connection provides a more rigorous theoretical foundation to our model for DF correction. We defer the full derivation to Appendix~\ref{AppendixC}, where we start from the collisional Boltzmann equation and summarize the steps leading to the Fokker-Planck formulation, ultimately showing how our DF correction emerges naturally as the first-order term in the diffusion equation, as previously shown by \cite{rosenbluth_fokker-planck_1957}, \cite{ipser_numerical_1977} and \cite{binney_galactic_2011} .

\subsection{Simulations set up} \label{subsect:simu_setup}

We generate several realizations having different resolutions of an NFW halo using the {\tt MAKEGALAXY} code (\citealt{springel_1999}), selecting an NFW DM density profile with a concentration parameter of $c = 4.38$ and virial mass $ \rm M_{200}=10^{13}$~M$_\odot$. This choices, consistent with \cite{genina_calibrated_2024}, ensure that for an orbiting BH of mass $\rm 10^9$~M$_\odot$ with initial halocentric distance $r=20  \ \mathrm{kpc} \simeq 1/4 \  \rm r_s$, the significantly larger mass of the host halo minimizes perturbations to the halo density and velocity distributions, preventing artificial alterations in its structure. The halo center position is calculated using the shrinking sphere algorithm developed by \cite{power_inner_2003}.

To explore the effects of DF on the BH in a more complex system, we also carry out an additional set of simulations at varying resolutions, incorporating a central stellar bulge core modeled with a Hernquist profile (see Sect. \ref{subsect: stellarbulge}). The bulge is characterized by a mass of $ \rm M_b=10^{11} \ M_\odot$ and scale radius $\rm r_s = 7$~kpc. 
The simulations are carried out using the {\texttt{OpenGADGET3}} code (\citealt{groth}, D24), which builds upon {\tt GADGET3}, which is in turn an improved extension of GADGET2 (\citealt{gadget2}). The code employs a hybrid Tree-Particle Mesh (TreePM) method to accurately resolve gravitational interactions across different spatial scales. In this approach, long-range gravitational forces are computed using a particle-mesh (PM) method, which efficiently solves Poisson’s equation on a grid, while short-range interactions are handled using a hierarchical tree algorithm, allowing for adaptive force calculations with high resolution in dense regions. When calculating the force softening between particles having different softening lengths, the code adopts the largest softening of the particles among the tree node. This detail is relevant for multi-component system and its effects will be analysed in Sect. \ref{Results.Stellarbulge}.


\subsubsection{List of simulations} \label{subsect:list_of_simu}
We introduce a BH having mass $\rm M_{BH}=10^{9} \ M_\odot$ at $\rm 20$ kpc from the halo center within the halo (eventually complemented with a central stellar bulge) configuration described in the previous sections. By default, BH particles are initialized on a circular orbit around the halo center, while we also investigate the case of an initially eccentric orbit in Appendix \ref{appendix_ecc} (see Fig. \ref{fig:ecc_numerical}).
Table \ref{tab:total_tab} summarizes the full set of simulations performed.
From left to right, we report in each column: the specific label assigned to each simulation, the BH to DM particle mass ratio $\rm M_{BH}/M_{DM}$, the BH to stellar mass ratio $\rm M_{BH}/M_*$ (where present), the number of particles used to sample the halo $\rm N_{Halo}$, the number of stellar particles used to sample the bulge $\rm N_{Bulge}$, the softening of the BH, DM and stellar particles, we indicate whether the simulations adopts the DF correction or not and finally what is the maximum impact parameter adopted in the DF correction. 

These simulations are divided into four main subgroups (separated by horizontal lines in Table \ref{tab:total_tab}), each designed to examine a specific aspect of the DF correction.

In the first set of simulations, we model the DM halo using progressively increasing particle numbers  ($\rm 10^5, 10^6, 10^7, 5 \cdot 10^7$). 
For each realization of the halo, we evolve the system both including and neglecting the effect of the DF correction of D24  (simulations labelled with $\rm *\_DF\_$ or $\rm *\_NODF\_$, respectively). We also adopt two different choices for the softening length:  following \citealt{power_inner_2003} (hereafter P03):
\begin{equation} \label{eqsoftP03}
    \epsilon_{P03}= \frac{4 R_{vir}}{\sqrt{N_{vir}}} 
\end{equation}
or according to \citealt{zhang_optimal_2019} (hereafter Z19):
\begin{equation} \label{eqsoftZ19}
    \epsilon_{Z19}= \frac{2 R_{vir}}{\sqrt{N_{vir}}}
\end{equation}
in simulations labelled with $\rm *\_p03$ or $\rm *\_z19$, respectively. 
Before initializing the BH orbit, we verify that adopting the smaller softening length $\epsilon_{\rm Z19}$, compared to the standard $\epsilon_{\rm P03}$, does not alter the density and velocity distributions of the system, which must remain consistent with the analytical predictions. This check is crucial to ensure that no spurious differences in the halo properties, possibly induced by different choices of the softening, affect the subsequent BH dynamics.
However, the choice of the softening length not only affects the performance of the N-body gravity solver but also the DF correction. Since we adopt $\rm b_{\rm max,c}=\epsilon_{BH}=\epsilon_{DM}$, by decreasing the softening we are also shrinking the region within which we compute the DF correction.

The purpose of this set of simulations is threefold: to assess the performance of the D24 model for DF correction across different resolutions, to test its numerical convergence against resolution, and to study the dependence on the choice of the softening length. 

A second set of simulations is devoted to the study of the same configuration but at lower resolution (beginning with $\rm LR\_$) when $\rm M_{BH} < M_{DM}$. Aiming to improve the DF prescription, we test the outcome when enlarging the impact parameter $ \rm b_{max, c}$ to two times the BH softening.

\begin{table*}[]
    \centering
    \caption{List of simulations.}

    \label{tab:total_tab}
\begin{tabular}{p{3cm } p{1.3cm} p{1.3cm} p{1cm} p{1.cm} p{1cm} p{1cm} p{1.5cm} p{1cm} p{1.2cm} }
    \hline
    \toprule
     Label & $\rm M_{BH}/M_{DM}$  & $\rm M_{BH}/M_{*}$  & $ \rm N_{Halo}$ & $ \rm N_{Bulge}$ & $\epsilon_{BH}$ (kpc) & $\epsilon_{DM}$ (kpc) & $\epsilon_{*}$ (kpc) & DF correction & $\rm b_{max,c}$ \\
     \hline
     \hline
      H1e5\_NODF\_p03 & $\rm 7 $ &  - & $10^5$ & - &  $4.43$ & $4.43$ & - & NO & -  \\
    H1e6\_NODF\_p03 & 70 &  - & $10^6$ & - &  $1.44$ & $1.44$ & - & NO & -  \\
    H1e7\_NODF\_p03 & $7  \cdot 10^2$ &  - & $10^7$ & - &  $0.44$ & $0.44$ & - & NO & -  \\
    H5e7\_NODF\_p03 & $\rm 3 \cdot 10^3$ &  - & $5 \cdot 10^7$ & - &  $0.22$ & $0.22$ & - & NO & -  \\ 
    H1e5\_NODF\_z19 & $\rm 7$ &  - & $10^5$ & - &  $2.22$ & $2.22$ & - & NO & -  \\
    H1e6\_NODF\_z19 & 70 &  - & $10^6$ & - &  $0.72$ & $0.72$ & - & NO & -  \\
    H1e7\_NODF\_z19 &  $7  \cdot 10^2$ &  - & $10^7$ & - &  $0.22$ & $0.22$ & - & NO & -  \\
    H5e7\_NODF\_z19 & $\rm 3 \cdot 10^3$  &  - & $5 \cdot 10^7$ & - &  $0.11$ & $0.11$ & - & NO & -  \\
          H1e5\_DF\_p03 & $\rm 7$ &  - & $10^5$ & - &  $4.43$ & $4.43$ & - & YES & $\epsilon_{BH}$  \\
    H1e6\_DF\_p03 & 70 &  - & $10^6$ & - &  $1.44$ & $1.44$ & - & YES & $\epsilon_{BH}$  \\
    H1e7\_DF\_p03 &  $7  \cdot 10^2$ &  - & $10^7$ & - &  $0.44$ & $0.44$ & - & YES & $\epsilon_{BH}$ \\
    H5e7\_DF\_p03 & $\rm 3 \cdot 10^3$  &  - & $5 \cdot 10^7$ & - &  $0.22$ & $0.22$ & - & YES & $\epsilon_{BH}$ \\ 
    H1e5\_DF\_z19 & $\rm 7$ &  - & $10^5$ & - &  $2.22$ & $2.22$ & - & YES & $\epsilon_{BH}$ \\
    H1e6\_DF\_z19 & 70  &  - & $10^6$ & - &  $0.72$ & $0.72$ & - & YES & $\epsilon_{BH}$  \\
    H1e7\_DF\_z19 &  $7  \cdot 10^2$ &  - & $10^7$ & - &  $0.22$ & $0.22$ & - & YES & $\epsilon_{BH}$ \\
    H5e7\_DF\_z19 & $\rm 3 \cdot 10^3$  &  - & $5 \cdot 10^7$ & - &  $0.11$ & $0.11$ & - & YES & $\epsilon_{BH}$ \\
    \hline
    LR\_H1e4\_NODF & 0.7&  - & $10^4$ & - &  $14$ & $14$ & - & NO & - \\
    LR\_H1e4\_DF & 0.7 &  - & $10^4$ & - &  $14$ & $14$ & - & YES & $\epsilon_{BH}$  \\
    LR\_H1e4\_DF\_2 & 0.7 &  - & $10^4$ & - &  $14$ & $14$ & - & YES & $2\epsilon_{BH}$  \\
    
    LR\_H7e3\_NODF & 0.5 &  - & $7 \cdot 10^3$ & - &  $19$ & $19$ & - & NO & -  \\
   
    LR\_H7e3\_DF\ & 0.5 &  - & $7 \cdot 10^3$ & - &  $16$ & $16$ & - & YES & $\epsilon_{BH}$  \\
    
    LR\_H7e3\_DF\_2 & 0.5 &  - & $7 \cdot 10^3$ & - &  $16$ & $16$ & - & YES & $2\epsilon_{BH}$  \\
        
        LR\_H5e3\_NODF & 0.35 &  - & $5 \cdot 10^3$ & - &  $19$ & $19$ & - & NO & -  \\
   
    LR\_H5e3\_DF\ & 0.35 &  - & $5 \cdot 10^3$ & - &  $19$ & $19$ & - & YES & $\epsilon_{BH}$  \\
    
    LR\_H5e3\_DF\_2 & 0.35 &  - & $5 \cdot 10^3$ & - &  $19$ & $19$ & - & YES & $2\epsilon_{BH}$  \\

    \hline
    B1e5\_NODF\_$\epsilon_{DM}$ & $7$ &  $1.4 \cdot 10^2 $ & $10^5$ & $2\cdot 10^4$  &  $4.43$ & $4.43$ & $1.63$ & NO & -  \\
    B1e6\_NODF\_$\epsilon_{DM}$ & 70 &  $1.4 \cdot 10^3 $ & $10^6$ & $2\cdot 10^5$  &  $1.44$ & $1.44$ & $0.51$ & NO & -  \\
    B1e7\_NODF\_$\epsilon_{DM}$ & $7  \cdot 10^2$ &  $1.4 \cdot 10^4 $ & $10^7$ & $2\cdot 10^6$  &  $0.44$ & $0.44$ & 0.16 & NO & -  \\

    B1e5\_NODF\_$\epsilon_{*}$ & 7 &  $1.4 \cdot 10^2 $ & $10^5$ & $2\cdot 10^4$  &  $1.63$ & $4.43$ & $1.63$ & NO & -  \\
    B1e6\_NODF\_$\epsilon_{*}$ & 70 &  $1.4 \cdot 10^3 $ & $10^6$ & $2\cdot 10^5$  &  $0.51$ & $1.44$ & $0.51$ & NO & -  \\
    B1e7\_NODF\_$\epsilon_{*}$ &  $7  \cdot 10^2$ &  $1.4 \cdot 10^4 $ & $10^7$ & $2\cdot 10^6$  &  $0.16$ & $0.44$ & $0.16$ & NO & -  \\
    
    B1e5\_DF\_$\epsilon_{DM}$ & 7 &  $1.4 \cdot 10^2 $ & $10^5$ & $2\cdot 10^4$  &  $4.43$ & $4.43$ & $1.63$ & YES & $\epsilon_{BH}$  \\
    B1e6\_DF\_$\epsilon_{DM}$ & 70 &  $1.4 \cdot 10^3 $ & $10^6$ & $2\cdot 10^5$  &  $1.44$ & $1.44$ & $0.51$ & YES & $\epsilon_{BH}$  \\
    B1e7\_DF\_$\epsilon_{DM}$ &  $7  \cdot 10^2$ &  $1.4 \cdot 10^4 $ & $10^7$ & $2\cdot 10^6$  &  $0.44$ & $0.44$ & $0.16$ & YES &  $\epsilon_{BH}$ \\

    B1e5\_DF\_$\epsilon_{*}$ & 7 &  $1.4 \cdot 10^2 $ & $10^5$ & $2\cdot 10^4$  &  $1.63$ & $4.43$ & $1.63$ & YES & $\epsilon_{BH}$  \\
    B1e6\_DF\_$\epsilon_{*}$ & 70 &  $1.4 \cdot 10^3 $ & $10^6$ & $2\cdot 10^5$  &  $0.51$ & $1.44$ & $0.51$ & YES & $\epsilon_{BH}$  \\
    B1e7\_DF\_$\epsilon_{*}$ &  $7  \cdot 10^2$ &  $1.4 \cdot 10^4 $ & $10^7$ & $2\cdot 10^6$  &  $0.16$ & $0.44$ & $0.16$ & YES &  $\epsilon_{BH}$ \\
    \hline

    LR\_B1e4\_NODF\_$\epsilon_{DM}$ & 0.7 &  14 & $10^4$ & $2\cdot 10^3$ &  $14$ & $14$ & $5.15$ & NO & -  \\

    LR\_B1e4\_NODF\_$\epsilon_{*}$ & 0.7 &  14 & $10^4$ & $2\cdot 10^3$  &  $5.15$ & $14$ & $5.15$ & NO & -  \\

    LR\_B1e4\_DF\_$\epsilon_{DM}$ & $\rm 10^{13}$ &  14 & $10^4$ & $2\cdot 10^3$  &  $14$ & $14$ & $5.15$ & YES & $\epsilon_{BH}$  \\

    LR\_B1e4\_DF\_$\epsilon_{*}$ & 0.7 &  14 & $10^4$ & $2\cdot 10^3$  &  $5.15$ & $14$ & $5.15$ & YES &  $\epsilon_{BH}$  \\
        
    LR\_B1e4\_DF\_2 & 0.7 &  14 & $10^4$ & $2\cdot 10^3$  &  $5.15$ & $14$ & $5.15$ & YES &  $2\epsilon_{BH}$  \\
    LR\_B1e4\_DF\_4 & 0.7 &  14 & $10^4$ & $2\cdot 10^3$  &  $5.15$ & $14$ & $5.15$ & YES &  $4\epsilon_{BH}$  \\

\hline\hline
\end{tabular} \tablefoot{The table presents, from left to right: the simulation label, the mass ratio of the BH to DM particles, the BH to stellar mass ratio, the total number of particles sampling the halo and the bulge, the softening lengths assigned to the BH, DM, and stars, whether the simulation includes the DF correction, and the maximum impact parameter considered in the DF correction.}
\end{table*}

Finally, a third set of simulations is carried out also including a central stellar bulge. We sampled the DM halo with $10^5, 10^6, 10^7$ particles and the stellar bulge with $2 \cdot10^4, 2 \cdot10^5, 2 \cdot10^6$ particles, respectively. The softening length for the star particles is scaled using the relation: 
\begin{equation} \label{stellar_soft}
    \rm \epsilon_{*}=\epsilon_{DM} \left( \frac{M_{*}}{M_{DM}} \right)^{1/3}.
\end{equation}
 The BH is initially placed at 20 kpc from the center, corresponding to approximately three times the bulge scale radius (see Sect. \ref{subsect:simu_setup}). As a result, the BH is initially embedded in a DM-dominated environment before gradually sinking into the stellar-dominated core. This raises the question of whether the optimal softening choice for the BH should follow the DM or stellar softening. To address this, we compare two approaches: {\em (i)} assigning to the BH the same softening as the DM particles $\epsilon_{BH}=\epsilon_{DM}=\epsilon_{P}$, and {\em (ii)} assuming for the BH the same softening of the star particles, $\epsilon_{BH}=\epsilon_{*}$ based on Eq.\eqref{stellar_soft}. 

Both configurations are tested in simulations with and without the DF correction. In cases where DF correction is applied, we adopt $ \rm b_{max,c}=\epsilon_{BH}$ as a reference choice. 

Similarly to the second simulation set but this time including the stellar central bulge, we carry out an additional low-resolution, fourth set of simulations where the BH mass is smaller than the typical DM particle mass $\rm M_{BH} < M_{DM}$. In this setup, we explore both choices of BH softening and, again, we perform two additional tests with increased value of the maximum impact parameter: $ \rm b_{max,c}=2\epsilon_{*}$. It is worth noticing that, in all the simulations including the bulge, the BH has a larger mass than the stellar particles, as indicated by the third column in Tab. \ref{tab:total_tab}, even for these low-resolution tests.


\section{Results} \label{Sec:Results}
Throughout this section, we outline the results of the simulations listed in Tab. \ref{tab:total_tab}. We present the results of the different sets of simulations. In Sect. \ref{Results.DMhalo} we analyse the timescale for sinking of a BH infalling into a DM halo. Sect. \ref{Results.Stellarbulge} presents the same results but for a multi-component system composed of a DM halo provided with a central stellar bulge.

\subsection{Infalling BH in a DM halo} \label{Results.DMhalo}
\begin{figure*}
\centering
    \includegraphics[width=0.99\linewidth]{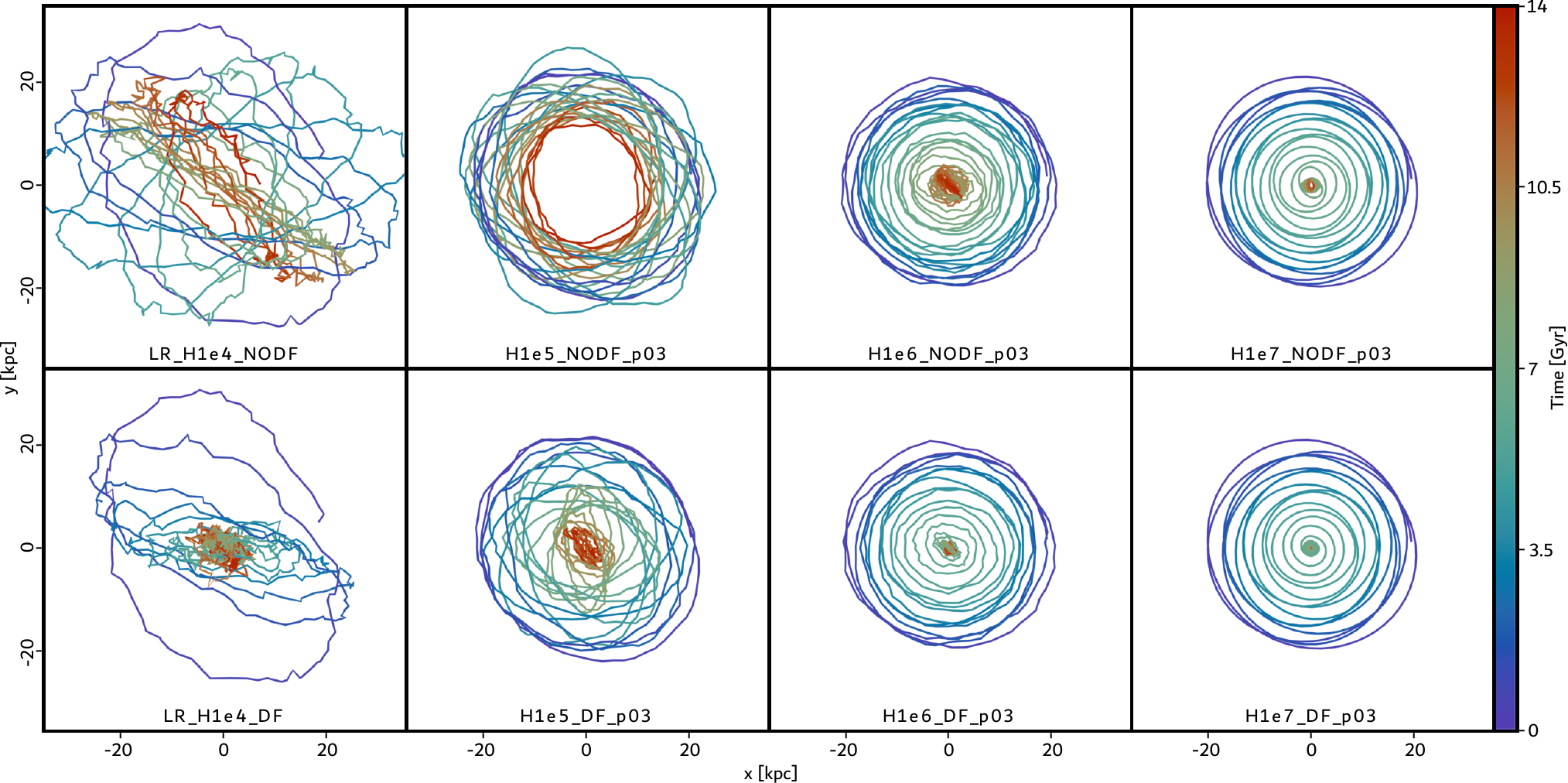}
    \caption{Projected trajectory of the BH in the DM halo on the initial orbital plane, for simulations at progressively increasing resolution, from left to right, as indicated by the labels (see Tab. \ref{tab:total_tab} for the description of the label of each simulation). 
    Trajectories are color-coded by time, as shown by the colormap. }
    \label{fig:numerical_trajectories}
\end{figure*}

For all the configurations described in this section, we place a BH of $\rm 10^9 M_\odot $ on a circular orbit at $20$ kpc from the DM halo center. Figure \ref{fig:numerical_trajectories} shows the trajectories of the BH  for simulations with progressively increasing resolution, displayed from left to right, as indicated by the labels below each panel (see Tab. \ref{tab:total_tab}). The orbits are projected in the initial orbital plane and color-coded by time, as shown in the colormap. We note that, regardless of whether the DF correction is applied, the BH develops an eccentric orbit, even if initially placed within a circular orbit, with the eccentricity decreasing as the resolution increases. At the two lowest resolutions, the effect of DF is clearly visible: the BH orbits decays to small halo-centric radii when DF is included, whereas the orbital decay is stalling without DF. In the following subsections, we quantify this behavior by analyzing the BH–halo center distance at different resolutions, with particular focus on the lowest-resolution cases (where $M_{\rm BH}< M_{\rm DM}$) discussed in Sect. \ref{DM.lowres}.  

\subsubsection{Reference resolutions}

\begin{figure*}
    \centering
    \includegraphics[width=0.9\linewidth]{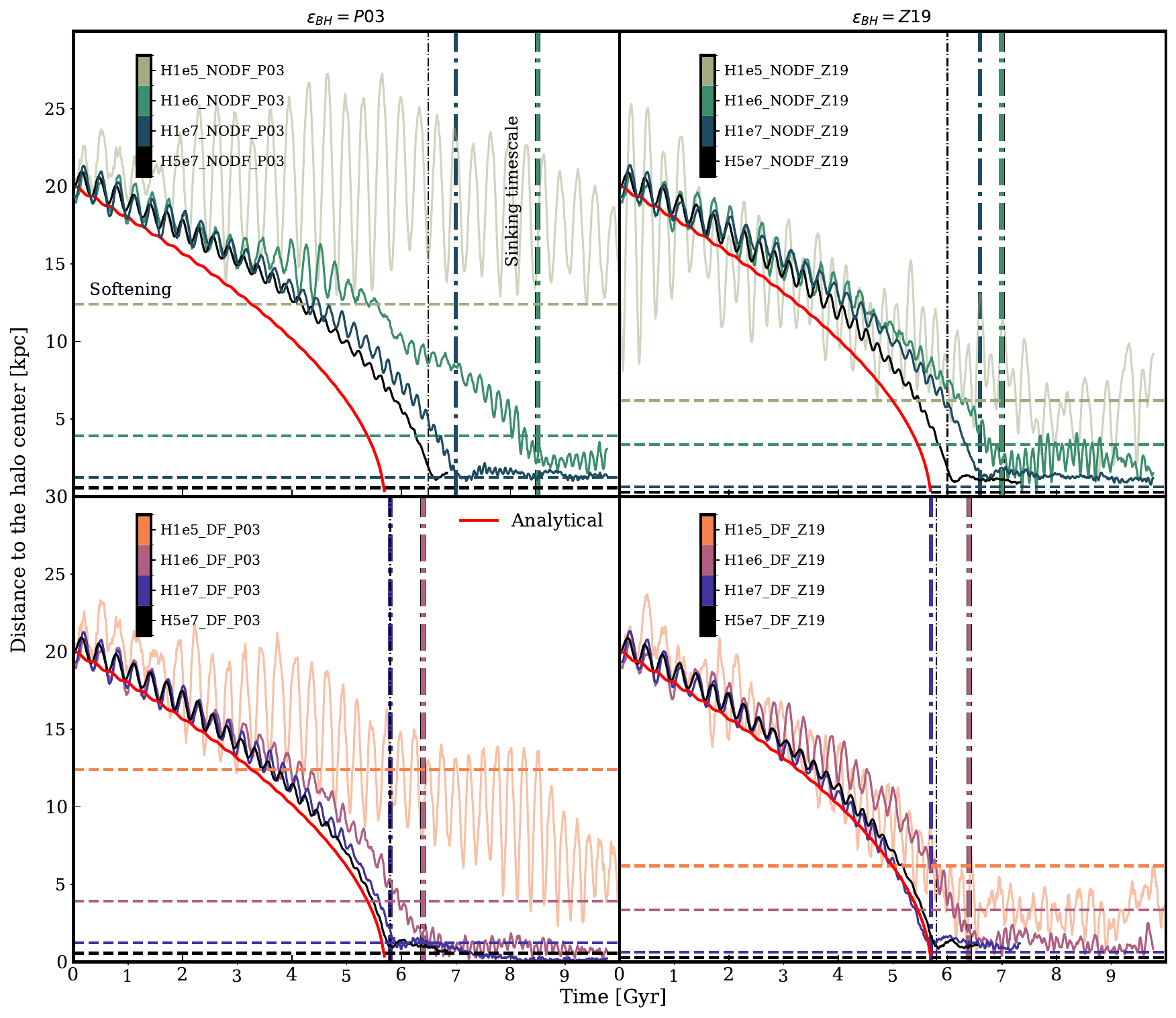} \caption{Sinking timescales for a BH initially seeded on a  circular orbits within a DM-only NFW halo at $20$ kpc from the halo center. The left and right columns show simulations with softenings derived from Eq.\eqref{eqsoftP03} and Eq.\eqref{eqsoftZ19}, respectively. Top plots refer to simulations without any DF correction, while the bottom plots include the DF correction according to D24. Lines are color-coded by resolution, with lighter to darker shades indicating increasing resolution (as shown in the colormap). Horizontal dotted lines represent the softening lengths, and vertical dash-dotted lines mark the sinking timescales for the three highest resolution runs. The red line shows the analytically predicted merging timescale using $\rm b_{max}=20$ kpc. }
    \label{fig:sinking_DMonly}
\end{figure*}

Figure \ref{fig:sinking_DMonly} shows the time evolution of the distance of the BH from the halo center in different simulations. The left and right columns show simulations where the particle softenings follow Eq.\eqref{eqsoftP03} and Eq.\eqref{eqsoftZ19}, respectively.
Top panels present the results of simulations without any sub-resolution prescriptions for BH dynamics, while the bottom plots show those where the BH dynamics includes the DF correction from D24. Curves are color-coded according to resolution, with darker shades representing higher resolution. Horizontal, dashed lines indicate the values of the softenings, color-coded according to the  resolution of the corresponding simulation.
We can visually identify the sinking timescale as the moment when the BH trajectory transitions from a steeper descent to a shallower, flatter curve. Vertical dash-dotted lines mark the sinking timescales hence identified for the three highest resolution simulations. The red line represents the predicted orbit retrieved from OTIS (Sect. \ref{subsect:otis}), using a maximum impact parameter $ \rm b_{\rm max}$ of 80 kpc, corresponding to the NFW halo scale radius.
In the simulations using the softening $\epsilon_{\rm BH} = \epsilon_{P03}$, we observe that at lowest resolution and without the DF correction (H1e5\_NODF\_P03), the BH does not sink toward the halo center in 10 Gyrs. However, when the DF correction is applied at the same resolution, we observe a faster sinking process where the BH crosses its softening scale at $ \sim 6$ Gyrs.
For higher-resolution simulations, the BH gradually sinks toward the center with or without the DF correction. 
For H1e6\_NODF\_p03 the sinking timescale is 8.5 Gyr, delayed by about 2 Gyr compared to analytical predictions. At higher resolutions, for  H5e7\_NODF\_p03, the timescale approaches 6~Gyr.
As we apply the DF correction,  H1e6\_DF\_p03, H1e7\_DF\_p03  and H5e7\_DF\_p03 predict sinking timescales in good agreement with theory, with a timescale for sinking  of 6.5~Gyr. Notably, H1e6\_DF\_p03 achieves the predicted timescale for sinking of H5e7\_NODF\_p03 but with a 50 times-lower resolution. 

Furthermore,  H1e7\_DF\_p03 and H5e7\_DF\_p03 converge to a sinking timescale of approximately 6~Gyr.
For simulations using $\rm \epsilon_{BH} = \epsilon_{Z19}$ and without the DF correction,  the sinking timescale is shorter than in the corresponding simulations with $\epsilon_{\rm BH} = \epsilon_{P03}$. 


When including the DF correction, the results for the three higher resolution simulations align closely with those adopting $\epsilon_{\rm BH}= \epsilon_{P03}$, thus indicating a weak sensitivity of our DF correction model on the softening choice at these resolutions.  On the other hand, H1e5\_DF\_z19 shows sinking timescales which are in agreement with theoretical expectations, despite the delay shown by the same simulation based on the P03 softening. 
In summary, the DF correction forces the BH to sink toward the halo center in the relatively low-resolution runs, while at higher resolution simulations show good agreement with theoretical predictions and resolution convergence. The additional convergence seen when reducing the softening length in line with Z19 indicates that: considering only particles within the $\epsilon_{\rm Z19}$ softening is sufficient to produce a DF correction that matches the analytical predictions; extending the correction region to $\epsilon_{\rm P03}$ for the higher resolution cases (where $\rm M_{BH}/M_{DM}>70$) does not significantly affect the results.

\subsubsection{Stability at lower resolutions}\label{DM.lowres}

All the simulation setups analyzed so far involved BH having a mass larger than the surrounding particles. However, in cosmological simulations, it is common to encounter situations where BHs have masses smaller than nearby DM and star particles, especially short after a BH is seeded. In this case, two-body encounters can spuriously heat the BH motion, resulting in an inaccurate representation of its trajectory. To mitigate possible shortcoming, simulations often couple the DF correction to a boosted dynamical mass unless, so that the mass of the BH entering in the computation of the gravitational force is set to be at least to a value comparable or exceeding the mass of the surrounding particles  (\citealt{chen}).
In this section, we describe how our particle-interaction-based approach to correct for the DF overcomes these limitations.
In Appendix \ref{AppendixC} we derive our DF correction from the Fokker-Planck equation, assuming that potential fluctuations affecting a mass $\rm M_{BH}$ arise from two-body interactions with a population of particles of mass $ \rm m_j$. 
In the formulation of D24, $ \rm \epsilon_{BH}$ defines the maximum impact parameter and the density surrounding the BH is assumed to be homogeneous.
While this approach does not impose any restrictions on the minimum allowed mass ratio $\rm M_{BH}/m_j$, it still struggles to accurately reproduce the sinking timescale for BHs that are less massive than the nearby particles.
To address this limitation, we refined the model by relaxing the assumption of a homogeneous surrounding medium. Additionally, we tested the correction by extending the maximum correction region beyond the softening length, i.e. using $ \rm b_{max, c} > \epsilon_{BH}$). In this section, we will show how these refinements improve the accuracy of the sinking timescale and ensure that the method remains robust across varying mass regimes, being also reliable when $\rm M_{BH}/m_j<1$. 

\begin{figure*}
    \centering
    \includegraphics[width=1\linewidth]{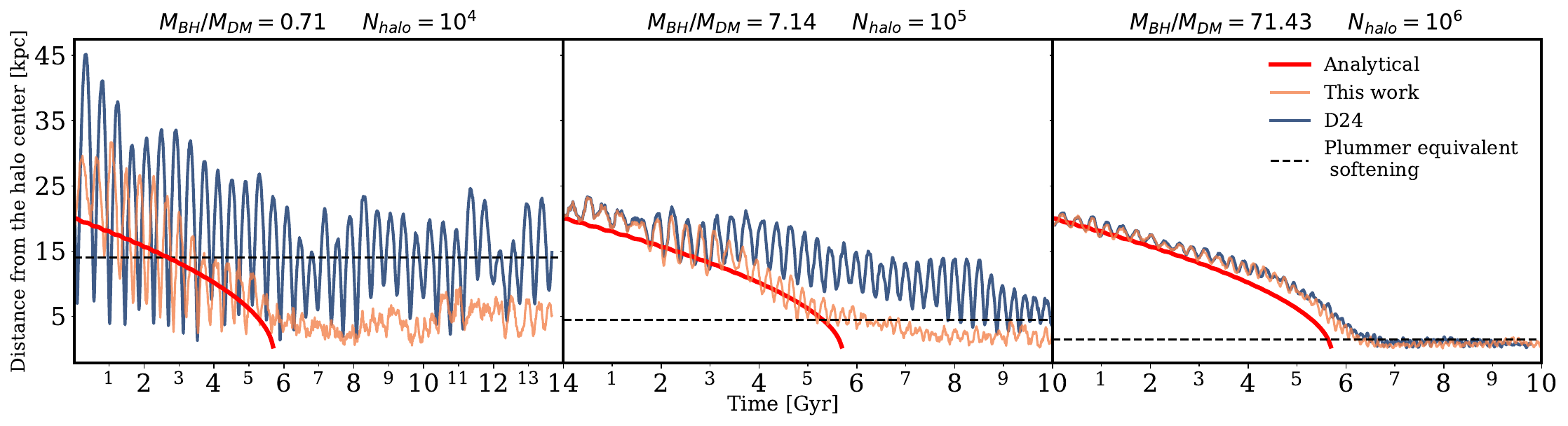} \quad
  
    \caption{Trajectories of a BH having $ \rm M_{BH} < M_{DM}$ placed on a circular orbit at $20$ kpc from the center of a DM halo for two different halo resolutions. From left to right, the  halo is sampled with $10^4$, $10^5$ and $10^6$ particles. The dashed black lines indicate the Plummer-equivalent softening length of each simulation.  Orange lines represent the BH trajectory when the DF correction is applied according to Eq.\eqref{df.loweres}, while the blue lines correspond to the original DF correction in Eq.(6) of D24, the red line shows the analytical prediction from OTIS (Sect. \ref{subsect:otis}). The ratio of the BH mass to the surrounding DM particle mass is shown at the top of each panel.}
    \label{fig:1e4_B1e9_nosoft_soft}
\end{figure*}

Starting from the cubic-spline kernel (\citealt{1985A&A...149..135M})
\begin{equation} \label{kernel}
W(r, \epsilon) = \frac{8}{\pi \epsilon^3} 
\begin{cases} 
1 - 6 \left(\frac{r}{\epsilon}\right)^2 + 6 \left(\frac{r}{\epsilon}\right)^3 &  ; \,0 \leq \frac{r}{\epsilon} \leq \frac{1}{2} \\ 
2 \left(1 - \frac{r}{\epsilon}\right)^3 & ; \, \frac{1}{2} < \frac{r}{\epsilon} \leq 1 \\ 
0 & ; \,\frac{r}{\epsilon} > 1\,,
\end{cases}
\end{equation}
where $\rm \epsilon$ is the softening length, the {\texttt{OpenGADGET3}} code derives the spline-softened gravitational force by taking the force from a point mass m to be the one
resulting from a density distribution $\rm \rho(r)=m W(r, \epsilon)$ (see \citealt{SpringelNaokiWhite}, Appendix A). Therefore, the local number density from the j-th particle placed at a position $\rm r_j$ in Eq.\eqref{general.df} is $ \rm \tilde{n}_j=W(r_j, \epsilon)$ and Eq.\eqref{general.df} becomes: 
\begin{eqnarray}  \label{df.loweres}
    \vec{a}_{\rm DF}= & &\rm \sum_j^{N(<b_{max, c})} -2 \pi G m_j (M_{BH}+ m_j){W(r_j, \epsilon)} \\ \nonumber 
     & & \times \rm \ln(1+\Lambda(m_j)^2) \frac{v_{m_j}-v_{BH}}{|v_{m_j}-v_{BH}|^3} \,.
\end{eqnarray} 
Figure \ref{fig:1e4_B1e9_nosoft_soft} shows the results of the application of Eq.\eqref{df.loweres} instead of Eq.\eqref{adf}, when $ \rm M_{BH} < M_{DM}$ (left plot) and 
 $ \rm M_{BH} > M_{DM}$ (central and right plots) for the usual DM halo characteristics and initial BH coordinates described in Sect. \ref{subsect:simu_setup} .
 
The dashed black line corresponds to the Plummer-equivalent softenings $\epsilon_{BH}=\epsilon_{P03}$ equal to $14.4$, $4,44$ and $1.44$ kpc from the left to the right panel. 
The orange curve in each plot indicates the trajectory of the BH when the DF correction is applied according to Eq.\eqref{df.loweres}. The blue curve corresponds to the original implementation of D24 (see Eq.\eqref{adf}). The red line indicates the analytical prediction according to the OTIS library.
At the top of each panel, we indicate the ratio of the BH mass to the mass of the surrounding DM particles. In the left plot, where the BH mass becomes smaller than that of nearby particles, using the DF correction adopted in D24, the BH cannot sink to the halo center but  oscillates with a maximum apocenter distance to the center of the halo higher than the Plummer-equivalent softening.
However, at this resolution, refining the description of the surrounding spatial distribution significantly improves the results: the BH sinks well below the Plummer-equivalent softening with timescales in agreement with the analytical results.
Even when $\rm M_{BH} > M_{DM}$ the refinement shows its success: the delayed sinking process predicted by the DF correction from D24 is now reduced and once again, the improved sinking timescale is in good agreement with the analytical predictions. 

Lastly, at higher resolution (right panel) the refinement has a negligible impact on the BH trajectory. The blue and orange trajectories are similar, indicating that the refinement can be applied with the same effectiveness as the original implementation in this regime.  

The second refinement that we test involves expanding the region over which the DF correction is applied. At lower resolutions, the assumption that DF effects are already well described by simulations at the resolved scales may no longer hold. To address this limitation, we explored the effect of assuming different values for the maximum impact parameter in the DF correction.

\begin{figure*}
    \includegraphics[width=1\linewidth]{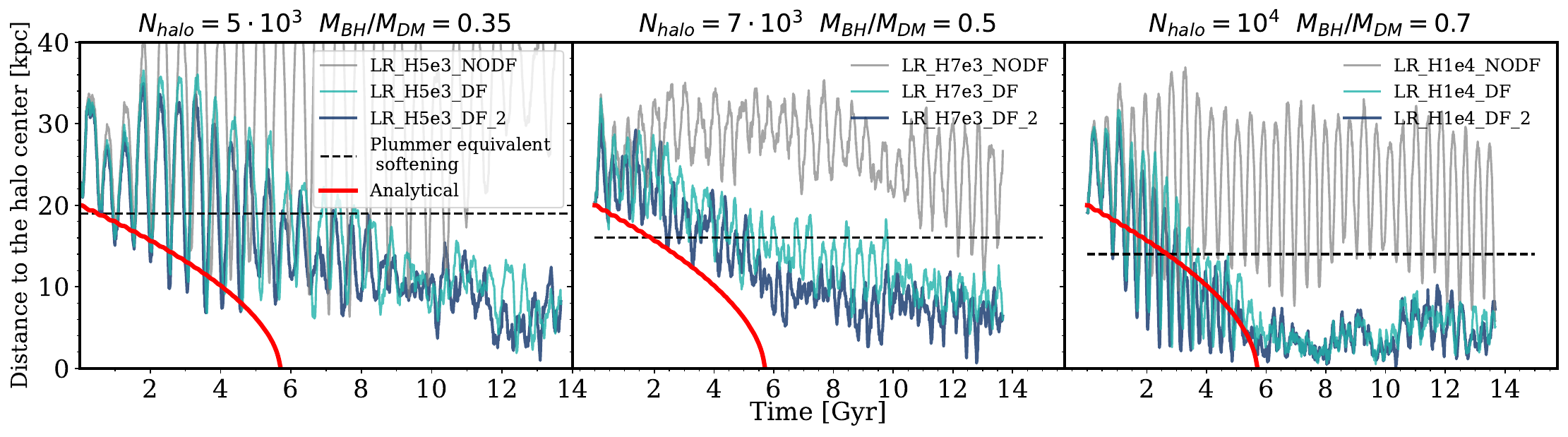}
 \caption{Sinking path of a BH of mass $\rm M_{BH}$ infalling in a DM halo sampled with DM particles of mass $\rm M_{DM}$ such that $\rm M_{BH}/M_{DM}< 1$. Each column refers to a different halo sampling, with increasing halo resolution from left to right. The grey curve corresponds to the simulation without the DF correction, whereas the blue curve refer to the simulation where the DF correction applied to the BH and computed according to Eq.\eqref{df.loweres} with $\rm b_{max,c}=2\epsilon_{BH}$ and $\rm b_{max,c}=\epsilon_{BH}$ (lighter curve). We report in red the analytical estimation from OTIS. The Plummer-equivalent softening in each simulation is marked with the horizontal dashed line. For the details of each simulation, refer to Tab. \ref{tab:total_tab}. }
 \label{fig:enlarge_region_1}
 \end{figure*}

Figure \ref{fig:enlarge_region_1} presents the results of these tests. From left to right, each column displays our findings when sampling the DM halo with $5 \times 10^3, 7 \times 10^3, 10^4 $ particles. In these cases, the BH is less massive than the DM particles, with corresponding mass ratios of $\rm M_{BH}/M_{DM}=0.35, 0.5, 0.7$. 
The grey curves represent results obtained without applying the DF correction, while the blue curves illustrate the trajectory of the BH when the DF correction is implemented according to Eq.\eqref{df.loweres}, with $\rm b_{max,c}= \epsilon_{BH}$ (lighter) or  $\rm b_{max,c}= 2\epsilon_{BH}$. The analytical solution from OTIS (Sect. \ref{subsect:otis}) is shown in red, and the dashed black line indicates the softening length of each simulation.

Across all the resolutions analyzed, the DF correction plays a crucial role in confining and driving the BH toward the center of the halo. Without the DF correction, the BH does not experience any orbital decay, whereas with the correction applied, it sinks well below the softening scale. Although this orbit shrinking occurs below the resolved scales, the process is captured with a timescale that is delayed 
when $\rm M_{BH}/M_{DM}=0.35, 0.5$ but aligns with the analytical prediction when $\rm M_{BH}/M_{DM}=0.7$. 
 However, we highlight that in all these low resolution tests, the BH exhibits considerable orbital scatter, and its trajectory is not circular.

 Lastly, increasing the size of the region within which we apply the DF correction does not lead to any significant improvement in the sinking timescale at any resolution. This suggests that particles located below the softening scale contribute most significantly to the DF correction. To validate this result, we further extended the corrective region to $\rm b_{max,c}=4\epsilon_{BH}$ and repeated the simulations with the same set-up presented in this section obtaining the same conclusions (see also Sect. \ref{Stellar.lowres}).

Based on the tests carried out in this section, we conclude that the DF correction introduced in D24 (see Eq.\eqref{adf}) is not adequate for accurately modeling the evolution of BHs when their mass is smaller than the mass of the surrounding particles. However, by introducing refinements that relax some of the underlying assumptions in the formula, we can significantly improve these results. Specifically, incorporating a spatially dependent phase-space density (Fig. \ref{fig:1e4_B1e9_nosoft_soft})  allows for a more accurate reproduction of the sinking process of BHs, even those lighter than the surrounding DM particles.

\subsection{Infalling BH in a composite DM and stellar bulge system} \label{Results.Stellarbulge}
In this section, we describe the results for the orbital evolution of a BH embedded in a composite system, made of a DM halo hosting an inner stellar bulge. The initial position of the BH is, as in the previous sections, $20 $ kpc from the halo center, and its initial velocity is the circular velocity at this halo-centric distance. Being the scale radius of the bulge $\rm r_{b}= 7.2$ kpc,  the BH is initially placed in a DM-dominated region, with its orbit lately decaying  into the stellar-dominated core.  As anticipated in Sect. \ref{subsect:simu_setup}, the {\texttt{OpenGADGET3}} code uses, for the interactions between particles with different softenings, the largest softening found for all the particles in each tree node. This is particularly relevant in a multi-softening system as the one presented in this section, where the BH experiences the transition from a DM-dominated to a stellar-dominated regime. To exploit the differences between different choices for $\rm \epsilon_{BH}$, we carried out simulations where the BH gravitational softening is set to be equal to both the DM, $\rm \epsilon_{BH}=\epsilon_{DM}$, and stellar particles one, $\rm \epsilon_{BH}=\epsilon_{*}$, with $\epsilon_{*}$ scales with resolution according to Eq.\eqref{stellar_soft}.
All the simulations are carried out both with and without the DF correction, using Eq.\eqref{df.loweres} with $\rm b_{max, c}=\epsilon_{BH}$. In Sect. \ref{Stellar.lowres} we investigate the effect of using a larger DF correction region.
\begin{figure*}
\centering
    \includegraphics[width=0.9\linewidth]{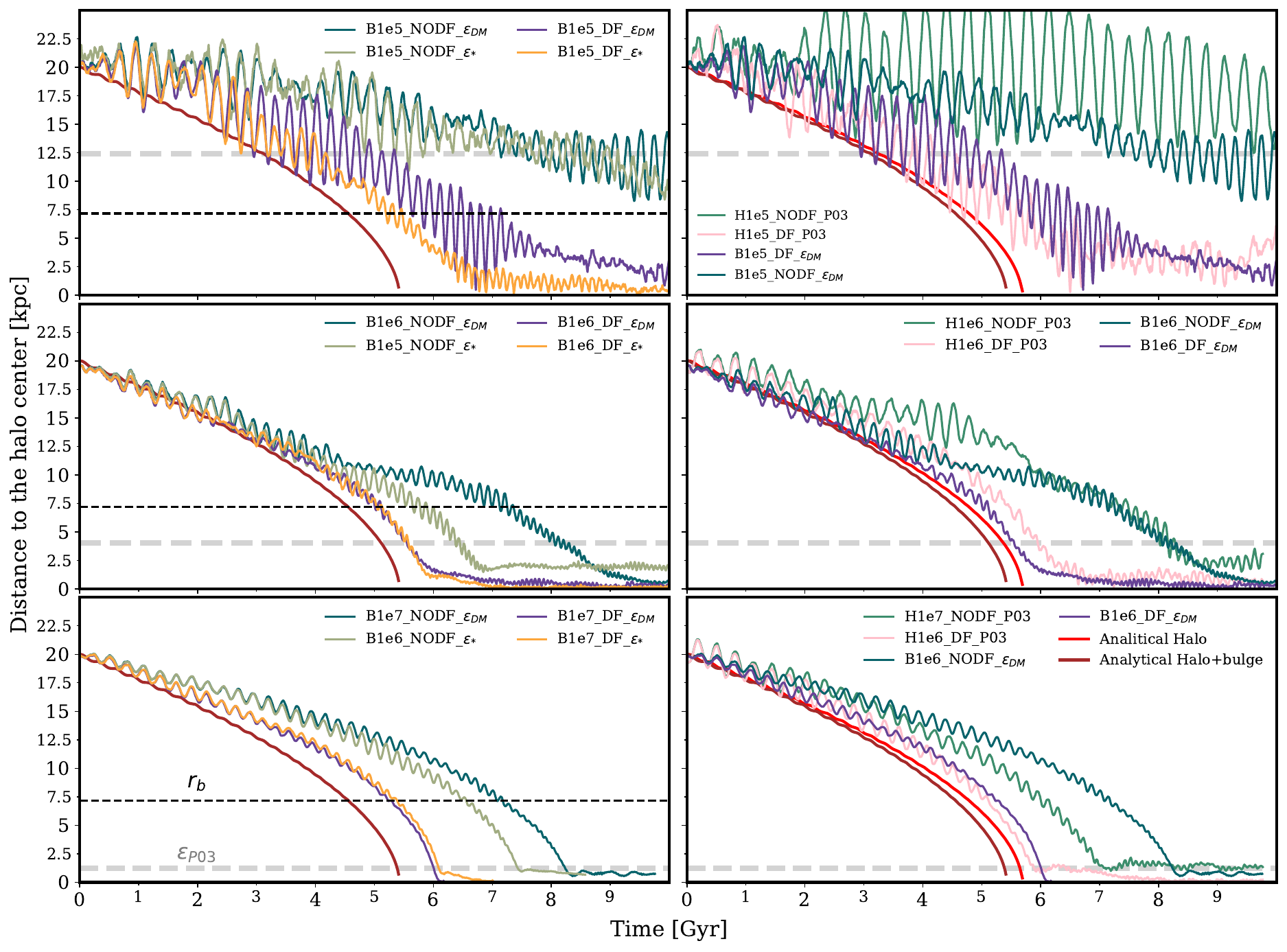}
    \caption{Comparison of the evolution of BH halo-centric distances during the BH sinking for different simulation setups. From the upper to the bottom row we show the results at increasing resolutions. Left column compares simulation using $ \rm \epsilon_{BH}= \epsilon_{DM}$ or $\rm \epsilon_{BH}=\epsilon_{*}$ for simulation with and without the DF correction. The right panels display the results for a BH embedded either in the NFW halo alone (as described in Sect. \ref{subsect:simu_setup}) or with the addition of a central stellar bulge. The red curves show the analytical predictions computed with OTIS, with the lighter curve corresponding to the DM-only case and the darker curve including the central bulge. Dashed horizontal lines indicate the Plummer-equivalent softenings used in the simulations, with Z19 softenings marked in black and P03 softenings in grey.}
    \label{fig:bulgeornot}
\end{figure*}
\subsubsection{Reference resolutions}
Fig. \ref{fig:bulgeornot} presents the results of these simulations, showing the distance to the system for different runs. The left panels compares simulations with and without the DF correction, as well as with different choices for the BH softening length: for $ \rm \epsilon_{BH}= \epsilon_{DM}$ or $\rm \epsilon_{BH}=\epsilon_{*}$. The meaning of each label is indicated in Tab. \ref{tab:total_tab}.  
From top to bottom, we show simulations at increasing resolution. Simulations at lower resolution and without the DF correction,   B1e5\_NODF\_$\epsilon_{DM}$ and B1e5\_NODF\_$\epsilon_{*}$, exhibit  the slowest orbital decay, crossing the BH softening length only after $\sim9$ Gyr, with no significant difference for different choices of $\epsilon_{BH}$. 
Increasing the number of particles sampling the halo and the bulge as for
 B1e6\_NODF\_$\epsilon_{DM}$,  the BH sinking timescale is noticeably delayed when approaching the bulge-dominated region.  Interestingly, this effect is less pronounced in B1e6\_NODF\_$\epsilon_{*}$. According to our choice for the force softening -- see Sect. \ref{subsect:simu_setup} -- in the DM-dominated region, the BH primarily interacts with the DM softening length, leading to a nearly identical trajectory for both softening choices.
However, as the BH approaches the bulge scale radius, the impact of softening differences becomes evident: larger softening values lead to longer sinking timescales, consistent with the findings in Sect. \ref{Results.DMhalo}.

Nevertheless, these differences largely vanish when comparing B1e5\_DF\_$\epsilon_{DM}$ and B1e5\_DF\_$\epsilon_{*}$, thus confirming that the introduction of the DF correction compensates for the different choices of softening. Simulations at higher resolution reinforce these trends, with only a mild deviation observed in B1e7\_NODF\_$\epsilon_{DM}$ when approaching $ \rm r_b$. However, also in this case,  the sinking timescale does not converge in simulations without the DF correction at different resolutions and different choices for the BH softening.

Since the presence of a central stellar component is expected to reduce the sinking timescale (as predicted by analytical estimates, see Fig. \ref{fig:otis_fig}), it is useful to compare the numerical results with and without the central bulge. The right column of Figure \ref{fig:bulgeornot} directly compares analogous simulations with and without the stellar bulge, using $ \rm \epsilon_{BH}=\epsilon_{DM}=\epsilon_{P03}$ in both cases, also comparing the versions with and without the DF correction. Interestingly, even when sampling the halo with $10^7$ particles - a configuration that successfully reproduced the results of the infall in the DM only case depicted in Fig. \ref{fig:sinking_DMonly} - the BH exhibits a longer sinking timescale in the presence of a central stellar bulge, an effect that is less pronounced when using the DF correction.  

\begin{figure}
    \centering
    \includegraphics[width=0.9\linewidth]{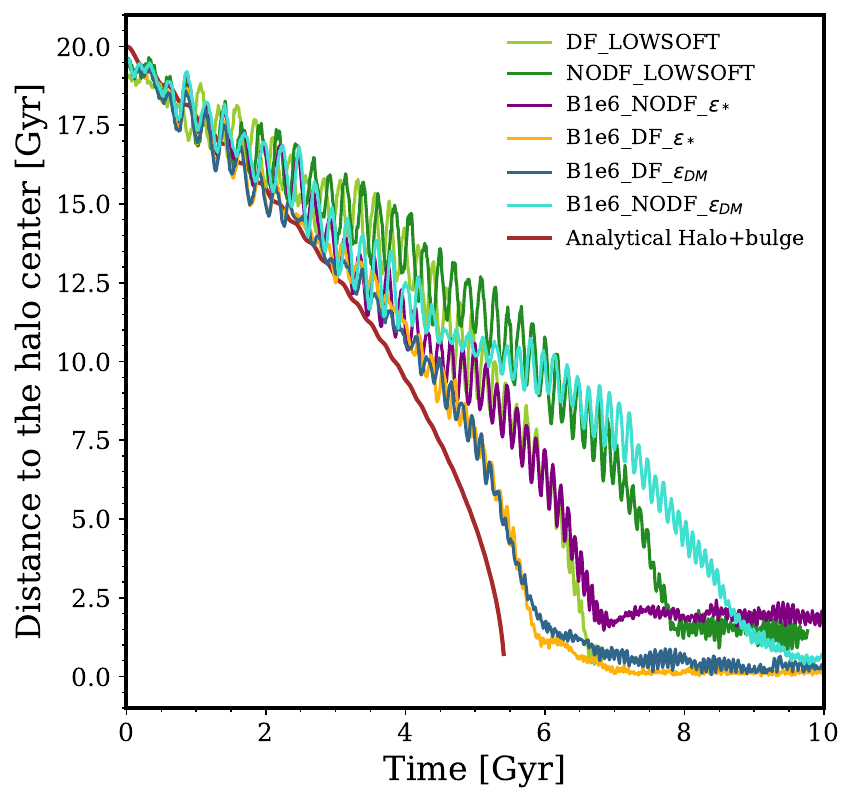}
    \caption{Evolution of the BH–halo center distance for simulations with the same mass resolution as $\rm B1e6\_DF\_\epsilon_*$ but using reduced gravitational softening lengths: $\epsilon_{\rm DM} = 0.72$, $\epsilon_{\rm BH} = \epsilon_* = 0.25$.  The green curves show the results from these simulations, with the darker line corresponding to the run without dynamical friction ($\rm NODF\_LOWSOFT$) and the lighter green line including the DF correction ($\rm DF\_LOWSOFT$). For reference, we plot in yellow and purple the outcome of $\rm B1e6\_DF\_\epsilon_*$ and $\rm B1e6\_NODF\_\epsilon_*$, while blue curves shows the results from  $\rm B1e6\_NODF\_\epsilon_{DM}$ (light) and $\rm B1e6\_DF\_\epsilon_{DM}$ (dark). The red line represents the analytical predictions  from OTIS.}
    \label{fig:lowsoft}
\end{figure}

The delay in the BH orbital decay is less pronounced when its gravitational softening length is set equal to that of the stellar particles. This suggests that this effect may be driven by the softened nature of the gravitational interactions. In order to verify this, we carried out two additional tests further reducing the softening lengths to the values $\epsilon_{\rm DM} = \epsilon_{\rm Z19}$ and $\epsilon_{\rm BH} = \epsilon_*$, where $\epsilon_*$ follows eq. \eqref{stellar_soft}. We show the results in Fig. \ref{fig:lowsoft}, which compares the BH–halo center distance in two simulations—one with DF, labelled as $\rm DF\_LOWSOFT$ and one without, namely $\rm NODF\_LOWSOFT$—with the same resolution as $\rm B1e6\_DF\_\epsilon_*$, but using the smaller softening lengths, so that $\epsilon_{\rm DM} = 0.72$, $\epsilon_{\rm BH} = \epsilon_* = 0.25$. 
We compare the results both with simulations at the same mass resolution but using a larger force softening and the analytical predictions from OTIS. 
Again, reducing the softening causes a significant delay in the orbital decay. 
This result highlights that simulations involving multiple particle species are particularly sensitive to softening choices. 
Deviations from analytic expectations can arise either due to softened gravitational interactions (for too large softenings) or from numerical heating caused by two-body encounters (for too small softenings). 
In both cases, we note that the application of the DF correction partially mitigates these effects.


\subsubsection{Stability at low resolution} \label{Stellar.lowres}
\begin{figure*}
    \includegraphics[width=1\linewidth]{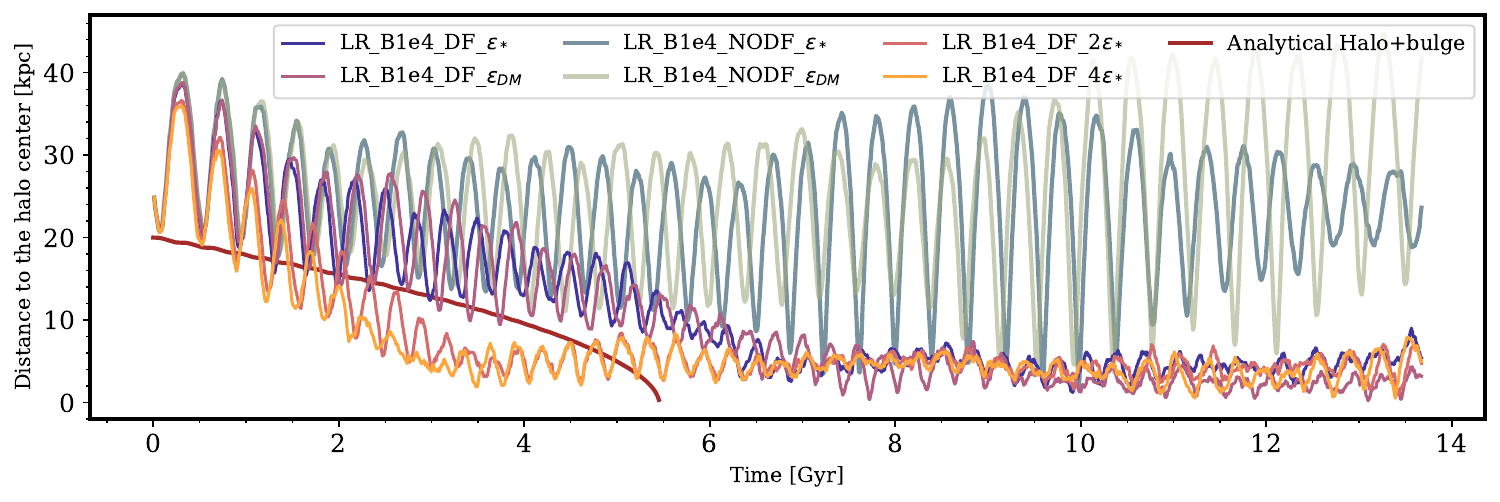}
    \caption{Evolution of the BH’s distance from the halo center in low-resolution simulations including both a DM halo and a stellar bulge. The figure compares different setups: 
     no DF correction (lighter and darker lines for different softening choices),  DF correction with $\rm b_{max,c}=\epsilon_{BH}$ and $\epsilon_{BH}= \epsilon_{DM}$ or $\epsilon_{BH}= \epsilon_{*}$  (lighter and darker purple lines), and DF correction with an extended impact parameter $\rm b_{max,c}=2\epsilon_{BH}, 4 \epsilon_{BH}$
  (lighter and darker orange lines). The brown curve represents the analytical sinking timescale from OTIS. }
    \label{fig:lr_bulge_sinking}
\end{figure*}
To extend our analysis to systems composed of particles with masses larger than the BH mass, we study the trajectory of the BH under these conditions, similar to the study presented in Section \ref{DM.lowres}. In the previous case, refining the D24 model by incorporating a kernel-dependent number density in Eq.\eqref{df.loweres} successfully reproduced the sinking process, even at low resolution, but we found no significant improvement when enlarging the corrective region. Given these findings, we conduct a comparable set of tests in the configuration that also includes a bulge component.
The results of these simulations are presented in Figure \ref{fig:lr_bulge_sinking}, where we examine the evolution of the BH distance from the halo center across different simulation settings. 

The figure compares cases without the DF correction either when $\rm \epsilon_{BH}=\epsilon_{DM}$ or $\rm \epsilon_{BH}=\epsilon_{*}$  (lighter and darker green curves), to those where the DF correction is applied using different impact parameters. We show the outcomes of simulations where $\rm b_{max,c}=\epsilon_{BH}$ in purple, where the darker line corresponds to $\rm \epsilon_{BH}=\epsilon_{DM}$ and the lighter one to $\rm \epsilon_{BH}=\epsilon_{*}$. We then increased the maximum impact parameter to $\rm b_{max, corr}=2\epsilon_{BH}= 2 \epsilon_*$ (dark orange) and $\rm b_{max, corr}=4\epsilon_{BH}= 4 \epsilon_*$ (light orange).

When the DF correction is not included, the BH orbit does not decay, regardless of the softening choice. However, once the correction is acting, the BH  sinks to the center, again showing no dependence on the adopted BH softening. Notably, increasing the maximum impact parameter results in an accelerated sinking process. Ultimately, enlarging beyond  $\rm \epsilon_{BH}$ the region within which the DF is corrected region does not improve the performances of the model in the DM halo-only case, while 
 leading to an overestimation of the DF in a multi-component system, eventually producing too short sinking timescales compared to analytical predictions. 

\subsubsection{Contributions to the DF acceleration} \label{subsect.relative}
In a system composed of both DM and stellar components, it is interesting to examine the relative importance of different terms in our model to correct the DF and, in particular, to assess the separate contributions of stars and and of DM to the overall DF correction.

\begin{figure}
    \includegraphics[width=0.9\linewidth]{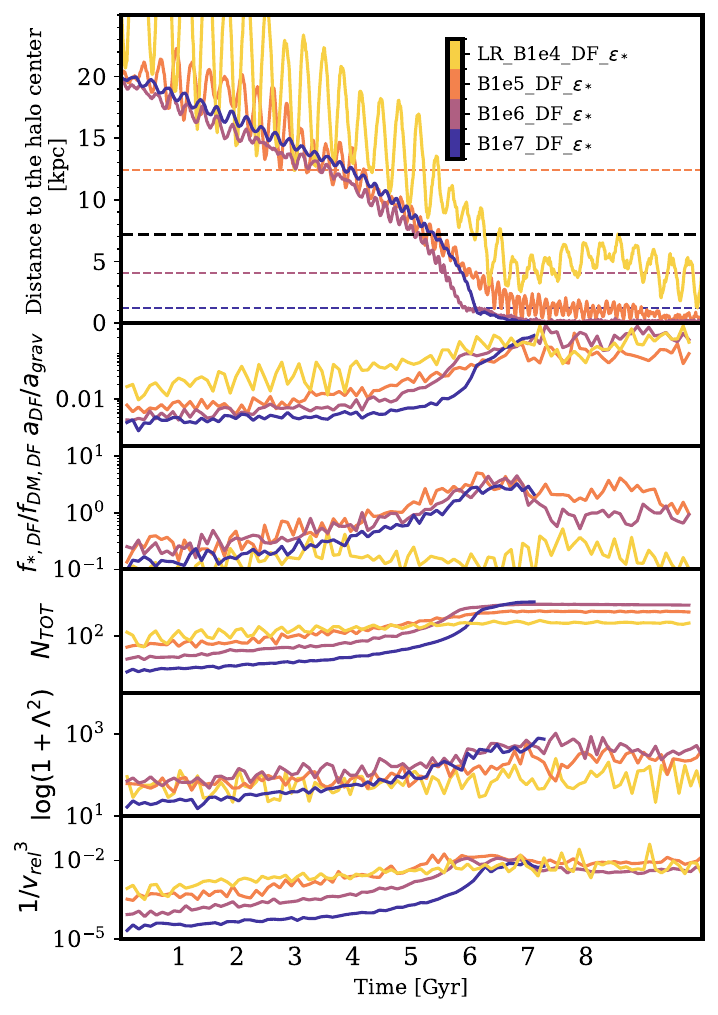}
    \caption{Breakdown of the different contributions to the DF correction term in Eq.\eqref{general.df} for simulations LR\_B1e4\_DF\_$\epsilon_{*}$, B1e5\_DF\_$\epsilon_{*}$, B1e6\_DF\_$\epsilon_{*}$, and B1e7\_DF\_$\epsilon_{*}$. From top to bottom, the panels show: BH distance to the halo center,  ratio of DF acceleration to gravitational acceleration, ratio of stellar to DM contributions to DF, total number of particles within $\epsilon_{BH}$, Coulomb logarithm term, and inverse velocity term of Eq.\eqref{inv.vel}.	 }
    \label{fig:relative_contributions}
\end{figure}

Fig. \ref{fig:relative_contributions} presents the analysis of these different components entering Eq.\eqref{general.df} for simulation LR\_B1e4\_DF\_$\epsilon_{*}$, B1e5\_DF\_$\epsilon_{*}$ and B1e6\_DF\_$\epsilon_{*}$,B1e7\_DF\_$\epsilon_{*}$.
The first panel from the top shows the evolution of the BH distance to the halo center, followed by the relative strength of the DF acceleration compared to the gravitational acceleration (second panel). The third panel illustrates the relative contributions of stars and DM  to the total DF correction, while the fourth panel shows the number of particles enclosed within $\rm \epsilon_{BH}$ (hence contributing to the DF correction). The last two panels report the value of $\log(1+\Lambda^2)$ and the term $\rm 1/v_{ rel}^3$: 
\begin{equation} \label{inv.vel}
    \rm \frac{1}{v_{rel}^3}= \sum_{i}^{N(<\epsilon_{BH})}\frac{1}{(v_{i}-v_{BH})^3}  \,.
\end{equation}
The horizontal dashed lines in the top panel mark the softening values color-coded according to the simulation they refer to.

As illustrated in the second row, a notable feature of these results is the enhanced DF correction at lower resolutions. With increasing resolution, DF is more accurately captured by the gravitational solver, reducing the need for an explicit DF correction, which consequently becomes progressively weaker. This effect is naturally accounted for in the prescription adopted, thus demonstrating that convergence to the analytical predictions can be achieved already at relatively low resolution.
 The relative contribution of the stellar component to DF also increases as the BH approaches the bulge-dominated region, except in the lowest-resolution simulation, where the softening is so high to already account for the entire stellar mass enclosed within the bulge from previous times in the simulation. This effect is further reflected in the number of particles within $\epsilon_{BH}$, which remains constant at the lowest resolution but increases for higher-resolution cases. Remarkably, the higher is the number of particles contributing to the DF correction, the lower is the relative velocity between the BH and the surrounding medium, leading to an increase of the velocity contribution in Eq.\eqref{df.loweres}.

 \section{Discussion}\label{Discussion}
In this work, we compared analytical predictions (Sect. \ref{Analytical}) and numerical simulations (Sect. \ref{Numerical}) of the sinking timescale for a BH infalling into a DM halo, both with and without a central stellar bulge. The analytical results were computed using OTIS python library to numerically solve the equations of motion of the BH (Sect. \ref{subsect:otis}).
Using the {\texttt{OpenGadget3}} code, we simulated the BH inspiralling onto the halo, exploring a wide range of resolutions and different softening choices. We applied a refined prescription to correct for unresolved DF, which extends the original D24 model, and extensively tested its impact on the BH dynamics.

\begin{figure*}
    \centering
    \includegraphics[width=0.9\linewidth]{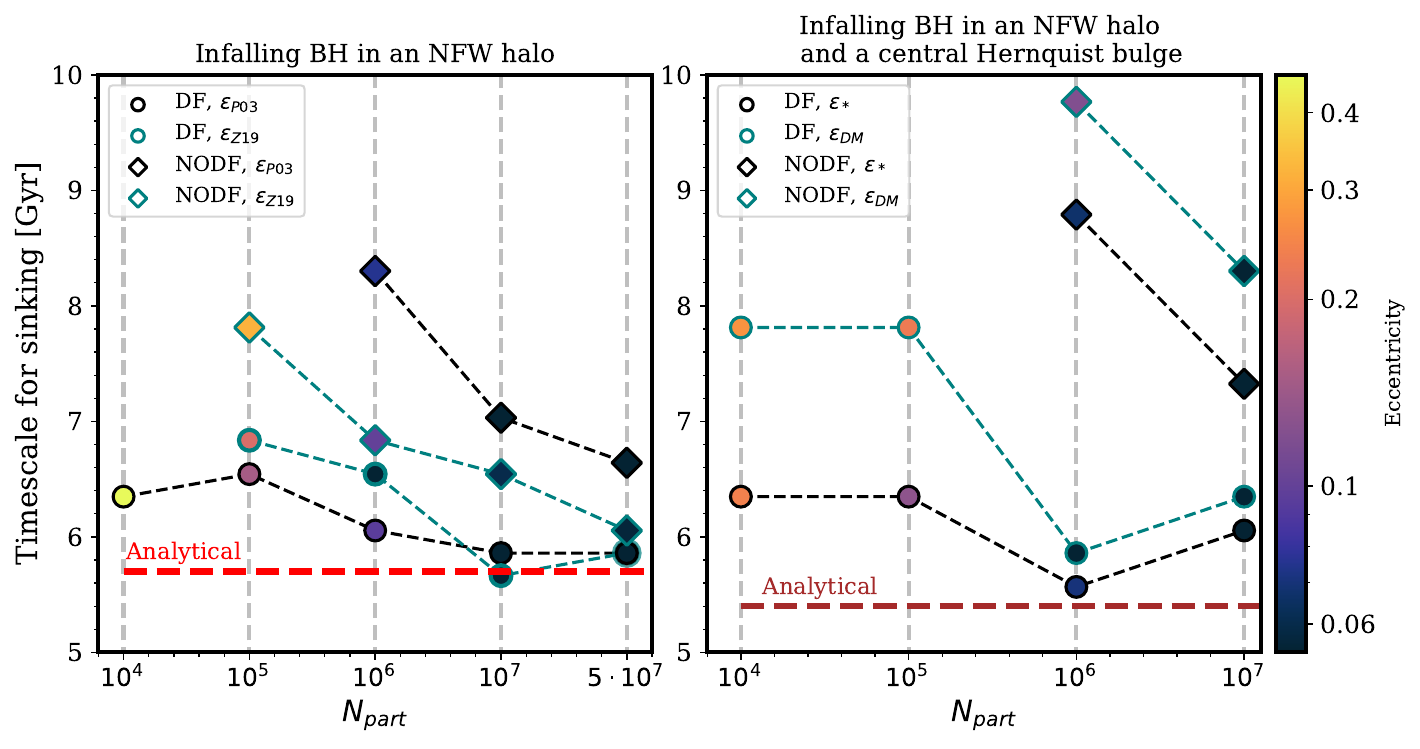}
    \caption{BH sinking timescale as a function of the total number of particles in the halo, for a pure DM NFW halo (left) and for a DM halo with a central Hernquist bulge (right). Diamonds indicate simulations without DF correction, circles indicate simulations with DF correction. The horizontal red dashed line shows the analytical sinking timescale from OTIS. In the left panel, teal-edged points, connected by teal dashed lines, adopt softening from eq. \eqref{eqsoftZ19}, black-edged points connected by black dashed lines are for the softening from eq. \eqref{eqsoftP03}. In the right panel, teal edges and teal dashed lines correspond to $\epsilon_{\rm BH} = \epsilon_{\rm DM}$, while black edges with black dashed lines correspond to $\epsilon_{\rm BH} = \epsilon_*$. The color of each point indicates the BH orbital eccentricity, as highlighted in the colorbar.}
    \label{fig:key_findings}
\end{figure*}
Figure \ref{fig:key_findings} summarizes some of our key findings of our analysis.
 In both panels, we plot the sinking timescales as a function of the total number of particles in the halo, when the BH is embedded in a pure DM NFW halo (left panel) or when a stellar Hernquist bulge is also included (right panel). Symbols represent different simulation setups: circles and diamonds indicate simulations with and without DF correction, respectively. The horizontal red dashed line marks the analytical sinking timescale predicted by OTIS. 
 In the right panel, teal-edged points correspond to simulations where the BH softening matches the DM particle softening ($\epsilon_{\rm BH} = \epsilon_{\rm DM}$), while black edges correspond to simulations where $\epsilon_{\rm BH} = \epsilon_{*}$.
The color of each point encodes the mean eccentricity of the BH during its inspiraling. Eccentricities are calculated by identifying local maxima and minima in the BH distance to the halo center (apocenter $r_{\rm apo}$ and pericenter $r_{\rm peri}$) and computing the eccentricity for each orbit as: \begin{equation} e = \frac{r_{\rm apo} - r_{\rm peri}}{r_{\rm apo} + r_{\rm peri}}. \end{equation} We then average the eccentricity over all orbits to obtain the color-coded value shown in the colorbar.

In the left panel, we observe that without the DF correction the BH experiences significantly longer sinking timescales. In low-resolution runs, they often even fail to sink, as discussed in Sect. \ref{DM.lowres}. Reducing the gravitational softening shortens the sinking time even without DF, but this effect remains insufficient to fully recover the analytical prediction. On the other hand, applying the DF correction produces sinking timescales that closely match the analytical predictions across all resolutions, and most importantly, reduces the sensitivity on the specific softening choice, as explained in Sect. \ref{Results.DMhalo}.

The inclusion of a stellar bulge, as depicted in the right plot, adds complexity to the BH sinking process. In principle, the higher central density contributed by the bulge, should accelerate the BH decay. However, our simulations reveal that numerical heating — especially at low resolution — counteracts this effect, thereby delaying the BH infall. Simulations without DF correction show pronounced deviations from analytical predictions, particularly at low resolutions. Quite remarkably, the application of the DF correction mitigates these numerical effects.
Moreover, Fig. \ref{fig:relative_contributions} illustrates that the DF correction is most significant in the dense central regions, where the relative velocity between the BH and surrounding particles decreases. In these conditions, the DF correction can contribute up to $\sim10\%$ of the total gravitational acceleration. As already pointed out by \citet{genina_calibrated_2024}, the dynamical heating of the BH induced by the numerical resolution not only affects the sinking timescale but also the orbital eccentricity. Figure \ref{fig:key_findings} shows that the lower the resolution the higher is the mean orbital eccentricity. Even initially placing the BH on a circular orbit, it acquires a mean eccentricity reaching up to $\sim 0.5$ at lower resolution. In this case, the addition of the DF correction only slightly mitigates this effect at low resolution. 

 \section{Conclusions}\label{conclusions}
In this paper, we presented results from the analysis of controlled numerical experiments aimed at describing the infall of a black hole (BH) into a dark matter (DM) halo, both with and without the inclusion of a central stellar bulge. The purpose of this analysis was to address the following key questions:
$(i)$ What is the impact of an additional dynamical friction (DF) correction to the BH gravitational acceleration in controlled numerical simulations?
$(ii)$ Does the sinking timescale obtained with the DF correction converge at progressively increasing resolution?
$(iii)$ How well do numerical results agree with analytical predictions at different resolutions?
$(iv)$ What is the impact of a stellar bulge on BH dynamics?
$(v)$ Can a DF correction effectively model BH sinking toward the halo center even in the unfavorable case when the BH mass is smaller than the surrounding particle masses?

To properly address these questions, we combined analytical calculations and numerical simulations at varying resolution. As for the analytical approach, outlined in Sect. \ref{Analytical}, we developed the OTIS library ( publicy available at: \url{https://github.com/alicedamiano5/OTIS.git }) to solve the equations of motion of an inspiralling BH in an NFW halo, optionally including a stellar bulge. As for the numerical approach, we used the N-body TreePM code {\texttt{OpenGadget3}} to simulate the sinking of a BH initially placed at $20$ kpc from the center of the halo. The DF correction applied consist of a refined version of the prescription introduced in \citet{damiano_dynamical_2024}, by incorporating a kernel-weighted estimate of the local density surrounding the BH, described in Sect. \ref{subsect:numerical_df} and \ref{DM.lowres}.
From this study, we draw the following main conclusions:
\begin{itemize}
     \item \textbf{Impact of the DF correction and numerical convergence}: The addition of the DF correction significantly reduces the BH sinking timescale compared to the cases of uncorrected DF. The results converge at increasing resolution, thus confirming the robustness of the implementation. See Sect. \ref{Results.DMhalo}, Fig. \ref{fig:sinking_DMonly}
\item \textbf{Comparison with analytical predictions}: Simulations with the DF correction closely match analytical expectations. In contrast, simulations without DF correction show that the BH fails to sink at low resolution or exhibits significant delays ranging from $2$ Gyr to $500$ Myr depending on resolution. See Sect. \ref{Results.DMhalo}, Fig. \ref{fig:sinking_DMonly}.

\item \textbf{Impact of the stellar bulge}: Differently from what expected from  analytical predictions, the inclusion of a stellar bulge delays the sinking process in numerical simulations due either softened interactions or numerical heating. However, the DF correction is able to mitigate this effect. We find that the stellar contribution to DF becomes significant—up to $10\%$—in the central stellar core. See Sect.\ref{Results.Stellarbulge} Fig. \ref{fig:bulgeornot} and Fig. \ref{fig:relative_contributions}.

\item \textbf{Stability at low resolution}: The refined DF model allows us to accurate tracking the BH inspiralling orbits even when the BH mass is lower than that of surrounding particles, a challenging regime for standard N-body simulations and typical of cosmological simulations. See Sect. \ref{DM.lowres}, \ref{Stellar.lowres}, Fig. \ref{fig:1e4_B1e9_nosoft_soft} and \ref{fig:lr_bulge_sinking}. 
\end{itemize}

As a general conclusion, our study demonstrates that the refined DF implementation—besides having a solid theoretical foundation (see Appendix \ref{AppendixC})—yields reliable results at high resolution and successfully models BH dynamics even in the critical low-mass regime. This has key implications in views of its application to cosmological simulations: reliable BH tracking is essential to accurately model BH-galaxy co-evolution, since BH motion affects gas accretion, the deposition of the ensuing AGN feedback energy and hence the subsequent evolution of the host galaxy. Moreover, our ability to model BH sinking in low-resolution regimes opens the door to carry out larger-volume simulations, which are crucial for building statistically robust predictions of gravitational wave event rates for current and future facilities such as PTA (e.g., \citealt{Hobbs_2010}) and LISA (e.g., \citealt{colpi2024lisadefinitionstudyreport}).

\begin{acknowledgements}
 AD gratefully acknowledges the University of Zurich for hosting part of this research, and thanks Pedro Capelo and Lucio Mayer for their valuable feedback and insightful discussions.
Simulations were performed at the CINECA Supercomputing Center (Bologna, Italy) with computing time allocated through national and CINECA-UNITS agreements, as well as at the INAF – Astronomical Observatory of Trieste PLEIADI cluster (\citealt{Bertocco.etal.2020, Taffoni.etal.2020}).
This paper has been supported by: the Fondazione ICSC, Spoke 3 Astrophysics and Cosmos Observations. National Recovery and Resilience Plan (Piano Nazionale di Ripresa e Resilienza, PNRR) Project ID CN\_00000013 "Italian Research Center on High-Performance Computing, Big Data and Quantum Computing" funded by MUR Missione 4 Componente 2 Investimento 1.4: Potenziamento strutture di ricerca e creazione di "campioni
nazionali di R\&S (M4C2-19 )" - Next Generation EU (NGEU); by the National Recovery and Resilience Plan (NRRP), Mission 4,
Component 2, Investment 1.1, Call for tender No. 1409 published on 14.9.2022 by the Italian Ministry of University and Research (MUR), funded by the European Union – NextGenerationEU– Project Title "Space-based cosmology with Euclid: the role of High-Performance Computing" – CUP J53D23019100001 - Grant Assignment Decree No. 962 adopted on 30/06/2023 by the Italian Ministry of Ministry of University and Research (MUR); by the INFN InDark Grant. This work was also supported by the Ministry of Education, Youth and Sports of the Czech Republic through the e-INFRA CZ (ID:90254).
\end{acknowledgements}

%
%
\bibliography{bibliography}

\begin{appendix}
\label{AppendixA}

\section{Infalling of a BH on an orbit with initial eccentricity}
The initial angular momentum of the BH decreases as we decrease its initial velocity, while keeping its initial position unchanged. According to Eq.\eqref{eccentricity}, this leads to an increase of the orbital eccentricity. In Sect. \ref{Analytical} we demonstrated that for eccentric orbits, the DF has two effects: it circularises the orbit as it approaches the center, and shortens sinking timescales as eccentricity rises. Here we examine how effectively the DF correction captures these effects across different resolution levels.

\begin{figure*}
    \centering
    \includegraphics[width=1\linewidth]{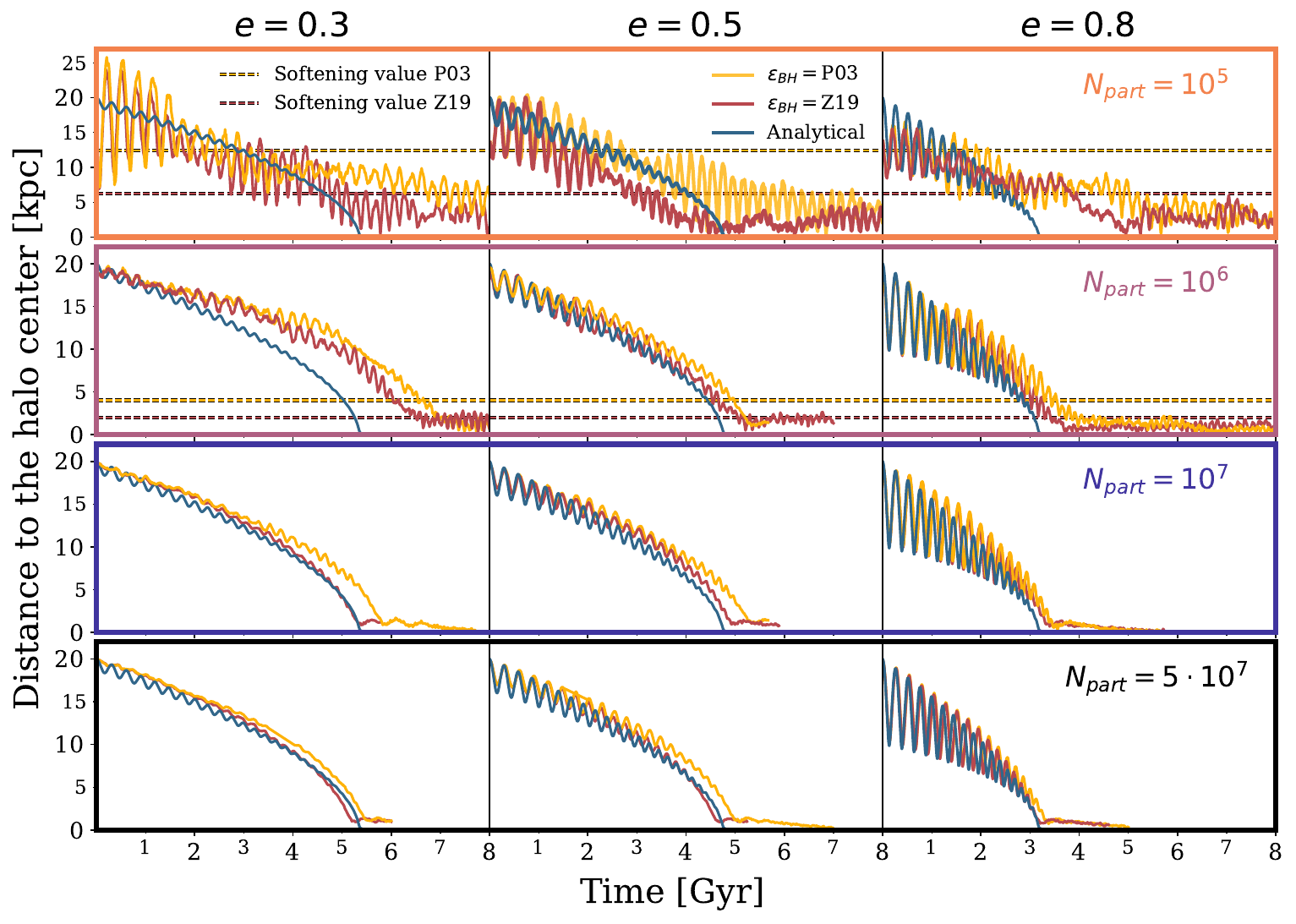}
    \caption{BH distance from the halo centre over time, starting from an initial position at 20 kpc, with orbital eccentricities $e=\rm 0.3, 0.5, 0.8$  shown in the first, second, and third columns, respectively. Each row represents a different halo resolution: $\rm N_{part}=10^5,10^6, 10^7, 5 \cdot 10^7$ from top to bottom. Yellow lines show results for simulations with P03 softening lengths, and red lines indicate Z19 softenings. Analytical predictions for each orbital eccentricity are represented as a blue line. Dotted lines mark softening values for each resolution (yellow edge for P03, red edge for Z19). All simulations apply the DF correction from D24.}
    \label{fig:ecc_numerical}
\end{figure*}

Figure \ref{fig:ecc_numerical} shows the BH distance from the halo center for an initial position of $20$~kpc with varying orbital eccentricities:  $e= \rm 0.3$, $e= \rm 0.5$, $e= \rm 0.8$ in the first, second and third column, respectively.
Each row shows the results of simulations with a different sampling of the same NFW halo, where resolution increases from top to bottom:  $N_{\rm part}=10^5$ in the first, $10^6$ in the second, $10^7$ in the third and $5 \cdot 10^7$ in the fourth row (the same halo realizations discussed in Sect. \ref{Results.DMhalo}). Yellow and red lines indicate results from simulations using softening length according to P03 (see Eq.\eqref{eqsoftP03}) or Z19 (see Eq.\eqref{eqsoftZ19}), respectively.  Analytical expectations from OTIS (see Sect. \ref{subsect:otis}) are shown as blue lines, obtained by setting $\rm b_{max}=r_s = 80 \ kpc$. Dotted lines represent the softening values of each resolution for P03 (yellow edge) and Z19 (red edge). All simulations incorporate the DF correction from D24 reported in Eq.\eqref{adf}. 
When sampling the halo with $10^5$ particles, variations on the BH orbit with the eccentricity are more evident before the BH crosses the softening length value. Within the first 2 Gyrs, the orbits are more eccentric compared to the analytical predictions for $e= \rm 0.3$ and less eccentric for $e= \rm 0.8$. 
Switching from P03 to Z19 softening further reduces the distance to the halo center due to the lower softening value of Z19. However, at low resolution, distinguishing different orbits according to their eccentricities is not possible; even though the BH is bound within the softening range around the halo center, no clear trend of the sinking timescale with increasing eccentricity is detectable.

At higher resolutions,  results from simulations align more closely with theoretical predictions and the sinking timescale becomes more sensitive to the initial orbital eccentricity. For simulations employing $\rm N_{part}=10^6$ and $10^7$ particles, numerical results increasingly approximate analytical expectations as eccentricity rises. Sinking timescales from simulation based on the  from P03 softenings are slightly delayed compared to those employing the Z19 softenings. 
\label{appendix_ecc}

\section{ A Fokker-Plank based derivation of the DF correction} \label{AppendixC}
In this appendix, we demonstrate that our model for the DF correction introduced in Eq.\eqref{num.general.df} can be recovered as the first-order diffusion coefficient of the Fokker–Planck equation to describe the collision term in the Boltzmann equation for a self-gravitating fluid. Our derivation follows primarily those presented by \cite{rosenbluth_fokker-planck_1957}, \cite{ipser_numerical_1977}, and \cite{binney_galactic_2008} (see in particular their Appendix L). For clarity, we start by briefly reviewing the derivation of the Fokker–Planck approximation, before demonstrating how the DF expression arises from its first-order term.
\subsection{The Fokker-Plank approximation}
Let us consider a stellar system described by the phase-space distribution function $f(\vec{x}, \vec{v},t)$
, where $\vec{x}$ and $\vec{v}$ denote the spatial and velocity vectors of the stars composing the system. Under the influence of an external force $\vec{F}$, and assuming each all stars have the same mass $m$, the Boltzmann equation reads
\begin{equation} \label{boltzmann} \frac{df}{dt} = \frac{\partial f}{\partial t} + v^{j} \frac{\partial f}{\partial x^j} + \frac{F^{j}}{m} \frac{\partial f}{\partial v^j} = \left( \frac{\partial f}{\partial t} \right)_{\text{coll}}\,. \end{equation}
Here, the index $j$ runs over the velocity components (i.e. $j = 1, 2, 3$) and the collision term on the right-hand side accounts for changes in the distribution function due to stellar encounters. This term is neglected under the assumption that all stars feel a smooth potential, so that the Boltzmann equation becomes a continuity equation in phase space. However, in a discretized system of gravitationally interacting particles,  collisions among pairs of particles add on top of the action of the smooth potential. The cumulative effect of such interactions generates the collisional effect of a dynamical friction (DF) force.

Under the assumption that the rate of change induced by these encounters is typically much smaller than that caused by the external  force, we can assume that within the small phase-space element $\Delta \vec{w} =  (\Delta \vec{x}, \Delta \vec{v})$ and over a short time $\Delta t$, the external field remains constant.

Let  $\vec{w}$ be the phase-space coordinate of a star at time $t$.
Let us also introduce the quantity $\psi(\vec{w, \Delta w})$ which is defined so that $\psi(\vec{w, \Delta w})\,d^6 (\Delta \vec{w})\Delta t$ is the probability that a star with coordinates $\vec{w}$ is scattered within the time interval $\Delta t$ to the coordinate $\vec{w+\Delta w}$ as a result of an encounter.
The phase-space distribution function   within this volume then changes due to particles  
that are scattered either into or outside the 6D volume element $\Delta \vec{w}$. The loss term (contribution from particles scattered out) is:
\begin{equation} \frac{\partial f}{\partial t} = -f(\vec{w}) \int d^6 \Delta \vec{w} \, \psi(\vec{w}, \Delta \vec{w}), \end{equation}
while the gain term (contribution from particles scattered in) is:
\begin{equation} \frac{\partial f}{\partial t} = \int d^6 \Delta \vec{w} \, \psi(\vec{w}-\Delta \vec{w}, \Delta \vec{w}) f(\vec{w} - \Delta \vec{w}). \end{equation}
Combining the two terms above, the collisional term in Eq.\eqref{boltzmann} can be cast in the form:
\begin{equation} \label{dfdt} \frac{df}{dt} = \int d^6 \Delta \vec{w} \left[ \psi(\vec{w} - \Delta \vec{w}, \Delta \vec{w}) f(\vec{w} - \Delta \vec{w}) - \psi(\vec{w}, \Delta \vec{w}) f(\vec{w}) \right]. 
\end{equation}

According to the Fokker-Plank approximation, particle-particle encounters are assumed to be "weak", meaning that $|\Delta \vec{w}|$ is small, so that we expand the first term in the square brackets of Eq.\eqref{dfdt} in a Taylor series up to the second order:
\begin{align} \psi(\vec{w} - \Delta \vec{w}) f(\vec{w} - \Delta \vec{w})&= \psi(\vec{w}, \Delta \vec{w}) f(\vec{w}) \nonumber \\ & -\sum_{\mu=1}^{6} \frac{\partial}{\partial w_\mu} \left[ \psi(\vec{w}, \Delta \vec{w}) f(\vec{w}) \right] \Delta w_\mu \nonumber \\ & + \frac{1}{2} \sum_{\mu,\nu=1}^{6} \frac{\partial^2}{\partial w_\mu \partial w_\nu} \left[ \psi(\vec{w}, \Delta \vec{w}) f(\vec{w}) \right] \Delta w_\mu \Delta w_\nu \,
\end{align}
where the indices $\mu$ and $\nu$ run over the phase-space components, i.e. both spatial and velocity coordinates.
Substituting this into Eq.\eqref{dfdt} we obtain
\begin{equation} \label{encounter.operator}  \frac{df}{dt} = -\sum_{\mu=1}^{6} \frac{\partial}{\partial w_\mu} \left[ D[\Delta w_\mu] f(\vec{w}) \right] + \frac{1}{2} \sum_{\mu,\nu=1}^{6} \frac{\partial^2}{\partial w_\mu \partial w_\nu} \left[ D[\Delta w_\mu \Delta w_\nu] f(\vec{w}) \right], \end{equation}
where we defined:
\begin{align} &D[\Delta w_\mu] = \int d^6 \Delta \vec{w} \, \psi(\vec{w}, \Delta \vec{w}) \Delta w_\mu, \label{diff1} \\& D[\Delta w_\mu \Delta w_\nu] = \int d^6 \Delta \vec{w} \, \psi(\vec{w}, \Delta \vec{w}) \Delta w_\mu \Delta w_\nu \label{diff2}. \end{align}
Eq. \eqref{diff1}, \eqref{diff2} are often referred to as the first- and second-order diffusion coefficients. Hereafter, we proceed under the assumption that the encounters are not only weak, but also local. In this case, the dominant contribution to the phase-space shift of a particle arises from interactions with nearby particles.
 If the encounters are local,  the  diffusion coefficients can be expressed purely as a function of the velocity shift $\Delta \vec{v}$, as in \cite{binney_galactic_2008}:
\begin{equation} \label{Dvi} D[\Delta v_i] = \int d^6 \Delta \vec{v} \, \psi(\vec{v}, \Delta \vec{v}) \Delta v_i, \end{equation} 

\begin{equation}  D[\Delta v_i \Delta v_j] = \int d^6 \Delta \vec{v} \, \psi(\vec{v}, \Delta \vec{v}) \Delta v_i  \Delta v_j. \end{equation}

In what follows, we focus on the first-order diffusion coefficient of Eq. \eqref{Dvi} which is the average change in velocity due to the encounters per unit time. Hence, this term captures the drift of the particle in velocity space and is related to the DF correction we applied. 
The second-order diffusion coefficient, on the other hand, describes the random scattering of the particle’s velocity, representing the diffusive spread caused by these interactions. We expect that this term becomes  important in the low-resolution regime, where stochastic fluctuations dominate. In the analysis presented in this paper we concentrated on the dynamical friction term, associated to the first-order diffusione coefficient. We defer to a future analysis a careful assessment of the ramdom-noise contribution associated to the second-order diffusion coefficient.
%

Before showing how the first-order coefficient relate to the DF acceleration, we review few basic results of two-body scattering.

\subsection{Two-body scattering} \label{two_body_scattering}
Let us assume to have two particles of mass $m$ and $M$ scattering with impact parameter $b$ and initial relative velocity $\vec{v}_0$. The particle of mass $M$, because of the interaction with $m$, experiences a deflection of its trajectory by an  angle $\theta$ which is related to the the impact parameter $b$ by the following expression (\citealt{binney_galactic_2008}): 
\begin{equation} \label{tantheta}
    \tan \left(\frac{\theta}{2}\right)=\frac{G(M+m)}{v_o^2 \ b}.
\end{equation}
Therefore, the impact parameter corresponding to a 90° deflection is \begin{equation} \label{b90}
    b_{90}=\frac{G(M+m)}{v_0^2}.
\end{equation}
The intial velocity $\vec{v}_0$ of particle $M$ changes as a consequence of the encounter. The change in velocity $\Delta \vec{v}$ can be decomposed in a parallel and an orthogonal component to the intial relative velocity $\vec{v}_0$, namely $\Delta \vec{v}_{\parallel}$ and $\Delta \vec{v}_{\perp}$. Again, following  \cite{binney_galactic_2008}, such components can be written as: 
\begin{equation} \label{deltav.perp}
    |\Delta v_{\perp}| = v_0 \frac{m}{m+M}\sin\theta =\frac{2mv_0}{M+m}\frac{b/b_{90}}{1+b^2/b_{90}^2}
\end{equation}
\begin{equation} \label{deltav.parallel}
    |\Delta v_{\parallel}| =   v_0 \frac{m}{m+M}(1-\cos \theta) =\frac{2mv_0}{M+m}\frac{1}{1+b^2/b_{90}^2}.
\end{equation}
When extending the analysis to a system populated by many particles, we can derive the differential scattering cross section for the encounters. Particles scattering from an annular region of area $d\sigma = 2\pi b db$ are deflected into an angular annulus of area $2\pi R \sin(\theta)R d\theta$ within the solid angle $d\Omega=2\pi \sin\theta d\theta$. Therefore, the differential scattering cross section $d \sigma/ d \Omega$, using Eq. \eqref{tantheta}, can be written as follows:
\begin{equation} 
\frac{d \sigma}{d \Omega} = \frac{b}{\sin (\theta)} \frac{db}{d \theta} =\frac{G(M+m)b}{2\sin(\theta) v_0^2 \sin^2 (\theta/2)}\,.
\end{equation}
Using now Eq.\eqref{tantheta} for $b$ and  $\sin(\theta)= 2 \sin(\theta/2) \cos(\theta/2)$ we obtain: 
\begin{equation} \label{diff.scatt.cross}
    \frac{d \sigma}{d\Omega} = \left[ \frac{G(M+m)}{2 v_0^2 \sin^2 (\theta/2)} \right]^2 = \left[ \frac{b_{90}}{2 \sin^2(\theta/2)} \right]^2
\end{equation}
where $b_{90}$ is defined in Eq.\eqref{b90}. 

 Based on the formalism introduced in this paragraph, we can now derive the DF correction term from the first-order diffusion coefficient of Eq.\eqref{Dvi}. 

\subsection{DF correction from the first order diffusion coefficient}

Let us consider different populations of particles with masses $m$, each described by a corresponding phase-space distribution function $f_{m}(\vec{v})$. The mean rate of change of velocity for a particle of mass $M$ moving at velocity $\vec{V}$, denoted as $D[\Delta V_i]$, is given by the cumulative effect of encounters with all $m$-mass particles. Following the formalism of \citet{rosenbluth_fokker-planck_1957}, Eq.\eqref{Dvi} can be rewritten using the differential scattering cross section as: 
\begin{equation} \label{dv1eq1}
D[\Delta V_i] = \sum_{m} \int d^3 v' \, f_{m}(\vec{v}') \int d\Omega  \frac{d\sigma(v_0, \theta)}{d\Omega}  v_0  \Delta v'_i, \end{equation} 
where $\vec{v_0} = \vec{v} - \vec{V}$ is the relative velocity between the background particle $m$ and the test particle $M$, and $d\Omega= 2\pi \sin\theta d\theta$ is the solid angle element. \citet{rosenbluth_fokker-planck_1957} solved the integral over the angular coordinate in Eq.\eqref{dv1eq1} by selecting a convenient reference frame. They defined an orthonormal basis $(\hat{e}'_1, \hat{e}'_2, \hat{e}'_3)$ such that $\vec{v}_0$ lies along $\hat{e}'_1$, i.e., $\vec{v}_0 \cdot \hat{e}'_1 = |\vec{v}_0|\equiv v_0$. Starting from any fixed orthonormal basis $(\hat{e}_1, \hat{e}_2, \hat{e}_3)$ we have then:
\begin{equation}  \label{change.basis}
\hat{e}_i \cdot \hat{e}'_1 = v_{0,i}/v_0 .
\end{equation}
where $v_{0,i}$ is the $i$-th component of $\vec{v_0}$ in the base $(\hat{e}_1, \hat{e}_2, \hat{e}_3)$.

Denoting by $\phi$ the angle between the orbital plane and the versor $ \hat{e'}_2 $, and following the sign convention of \citet{binney_galactic_1987} to ensure that the velocity change is positive when the force between the particles is attractive, as in the case of the gravitational force, we can express the change in velocity in the $ (\hat{e'}_1, \hat{e'}_2, \hat{e'}_3) $ reference frame as:
\begin{equation} \label{deltav_e'}
\Delta \vec{v} = \Delta v_{\parallel} \hat{e'}_1 + \Delta v_{\perp} \cos(\phi) \hat{e'}_2 - \Delta v_{\perp} \sin(\phi) \hat{e'}_3,
\end{equation}
where $\Delta v_{\parallel} $ and $ \Delta v_{\perp} $ are the parallel and perpendicular components of the velocity change relative to $ \vec{v}_0 $ (and hence to $ \hat{e'}_1 $), as introduced in Eqs.~\eqref{deltav.parallel} and \eqref{deltav.perp}. 
Furthermore, in the generic reference frame with basis vectors $(\hat{e}_1, \hat{e}_2, \hat{e}_3) $, the components $ \Delta v_i $ can be expressed as:
\begin{equation} \label{deltav_e'_e}
    \Delta v_i = (\Delta \vec{v} \cdot \hat{e}'_k) (\hat{e}_i \cdot \hat{e}'_k).
\end{equation}
Assuming that the \( m \)-populations are isotropically distributed in space, we can average over the angle $\phi $. Since $ \langle \cos \phi \rangle_\phi = \langle \sin \phi \rangle_\phi = 0 $, the second and third terms in Eq.~\eqref{deltav_e'} vanish upon averaging. Combining Eqs.~\eqref{deltav_e'} and \eqref{deltav_e'_e}, we then obtain:
\begin{equation}
\Delta v_i = \Delta v_{\parallel} (\hat{e}_i \cdot \hat{e}'_1),
\end{equation}
so that Eq. \eqref{dv1eq1} becomes
\begin{equation}
    D[\Delta V_i]= \sum_{m} \int dv' f_{m}(v') \int d \Omega \frac{d\sigma (v_0, \theta)}{d \Omega} v_0 \Delta v_\parallel (\hat{e_i} \cdot \hat{e'_1})\,.
\end{equation}
Using then Eq. \eqref{diff.scatt.cross} and \eqref{change.basis}, we can write the diffusion coefficient as 
\begin{eqnarray}
    D[\Delta V_i] = 2 \pi \sum_{m}&&\int d^3 v' f_{m}(v') \left[  \frac{b_{90}}{2}\right]^2 v_0 {v}_{0_i} \times \nonumber \\
    && \int_{\theta_{min}}^{\theta_{max}} \frac{\sin \theta}{\sin^4(\theta/2)} (1-\cos \theta) d\theta.
    \label{dv1eq2}
\end{eqnarray} 
 In principle, the integration should run over the total solid angle. However, we  introduced the integration extremes $\theta_{min}, \theta_{max}$ that, according to Eq. \eqref{tantheta} correspond to the  scattering angle from a maximum and a minimum impact parameter. We adopt $b_{\mathrm{min}} = b_{90}$ and $b_{\mathrm{max}} = \epsilon_{\mathrm{BH}}$. The parameter $\epsilon_{\mathrm{BH}}$ is the gravitational softening length of the BH and serves to account for unresolved, sub-resolution diffusion effects, as discussed in Sect. \ref{subsect:numerical_df}.
Focusing on the integral over the angular coordinate, it can be easily solved noticing that: 
$\sin^4 (\theta/2)=(1-\cos\theta)^2/4$ so that: 
\begin{eqnarray}
\int_{\theta_{min}}^{\theta_{max}} \frac{\sin \theta}{\sin^4(\theta/2)} (1-\cos \theta) d\theta &=& \int_{\theta_{min}}^{\theta_{max}} \frac{4\sin \theta}{1-\cos(\theta)}d\theta \nonumber \\
&=& 4\ln(1-\cos \theta)\big|_{\theta_{min}}^{\theta_{max}}
\label{integral.angle}
\end{eqnarray}
 Moreover, from Eq.\eqref{deltav.parallel} we have:
\begin{equation} \label{1costheta}
    1-\cos \theta = \frac{2}{1+b^2/b^2_{90}}
\end{equation}
Using the results of Eqs. \eqref{integral.angle} and \eqref{1costheta} and  the minimum impact parameter as $b_{min}=b_{90}$ defined in Eq. \eqref{b90} we get:
\begin{equation} \label{final.dv1}
    D[\Delta V_i]= 2 \pi \sum_{m}\int d^3 v' f_{m}(v') \frac{G^2 (M+m)m}{v_0^3}\ln \left(1+\frac{b_{\rm max}^2}{b^2_{90}} \right) v_{0_i}.
\end{equation}
 Let us now assume to have a single  population of "stellar" particles of mass $m_j$ within a spherical region of radius $\epsilon_{BH}$ whose phase-space density distribution is described by the expression
\begin{equation} \label{distr.func}
\rm f(\vec{v}) =  \sum_{j}  \tilde{n}_j \delta(\vec{V}- \vec{v_j})\,,
\end{equation}
where the index $j$ runs over all the particles in the population. In our description of the sub-resolution correction of the DF force, these are the "neighbors" of the BH particle, that lie within the BH smoothing length. 
Using Eq.\eqref{distr.func} for the phase-space distribution function and $b_{\rm max}= \epsilon_{\rm BH}$, Eq.\eqref{final.dv1} becomes: 
\begin{equation} \label{dv1.equivalent}
        D[\Delta V_i]= 2 \pi \sum_{j}  \tilde{n_j} \frac{G^2 (M+m_j)m_j}{|\vec{v}_j-\vec{V}|^3}\ln \left(1+\frac{(\vec{v}_j-\vec{V})^4 \epsilon_{\rm BH}^2}{[G(M+m_j)]^2} \right) ({v_{i,j}}-{V_i}).
\end{equation}
Following the definition adopted in D24, we write:
\begin{equation} \label{lambda.numerical}
\Lambda(m_j) = \frac{(\vec{v}j - \vec{V})^2 , \epsilon{\rm BH}}{G(M + m_j)},
\end{equation}
which allows us to recognize that Eq.\eqref{dv1.equivalent} is formally equivalent to the correction term introduced in Eq.\eqref{num.general.df}.

As a final remark, it is important to stress that the first-order diffusion coefficient can be interpreted as an acceleration. This ultimately justifies the fact that the above expression for the diffusion coefficient corresponds to the DF correction of Eq.~\eqref{num.general.df}. As pointed out when we introduced it in Eq.~\eqref{Dvi}, this coefficient represents the average rate of change of the velocity—already suggesting its physical meaning as an acceleration. However, \citet{rosenbluth_fokker-planck_1957} provided an elegant formulation that explicitly demonstrates how this drift term can be related to a force. Following their approach, we consider Eq.\eqref{final.dv1}, restricting for clarity to a single stellar population of mass $m$, so that the summation reduces to a single term and assuming that the minimum and maximum impact parameters are velocity-independent. Then,Eq.\eqref{final.dv1} becomes:
\begin{equation} \label{dvi.eq3}
D[\Delta v_i]= 2 \pi G^2 (M+m)m\ln \left(1+\frac{b_{\rm max}^2}{b^2_{90}} \right) \int \mathrm{d}^3 v' f_{m}(v') \frac{v_{0_i}}{v_0^3}.
\end{equation}

After defining one of the two "Rosenbluth potentials" as 
\begin{equation} \label{rosenbluth.potential}
h_{m}(v) = \int \mathrm{d}^3 v' f_{m}(v') \frac{1}{v_0},
\end{equation}
and owing to 
\begin{equation}
\frac{\partial}{\partial v_i} \left( \frac{1}{v_0} \right) = \frac{v_{0_i}}{v_0^3}\,,
\end{equation}
we can express the first-order diffusion coefficient in terms of a force $\vec{F}_{\rm drift}$: 
    \begin{equation} \label{deltav_withnabla}
    D[\Delta \vec{V}]= 2 \pi G^2 (M+m)m\ln \left(1+\frac{b_{\rm max}^2}{b^2_{90}} \right) \vec{\nabla_v} h_{m}(v) = \frac{\vec{F}_{ \rm drift}}{M}\,.
\end{equation}
 In the above equation, $\vec{F}_{\rm drift}$ is in fact an effective force in velocity space, since it is proportional to the gradient, with respect to the velocity coordinate, of a velocity-dependent potential. \cite{binney_galactic_2011} highlighted that the Rosenbluth potential $h_m(v)$ depends on velocity in exactly the same way as the gravitational potential depends on position, thus emphasising the analogy between the gravitational force in real space and the DF force in velocity space.
If $M=m$ we can describe the evolution of the phase-space distribution function $f_m$ by combining Eq. \eqref{boltzmann} and the first-order term appearing in Eq. \eqref{encounter.operator} to obtain a Boltzmann equation in which an additional "drift force term" appears, according to: 
\begin{equation}  \frac{df}{dt} = \frac{\partial f}{\partial t} + v^{j} \frac{\partial f}{\partial x^j} + \frac{F^{j}}{m} \frac{\partial f}{\partial v^j} = \frac{\partial}{\partial v^j}\left(\frac{F^j_{\rm drift}}{m}f\right).
\end{equation}

\end{appendix}

\end{document}